# *A Quantum Computer Foundation for the Standard Model and SuperString Theories*[*]

**By**

**Stephen Blaha**[**]

# ABSTRACT

1. SuperString Theory can naturally be based on a Quantum Computer foundation. This provides a totally new view of SuperString Theory.

2. The Standard Model of elementary particles can be viewed as defining a Quantum Computer Grammar and language.

3. A Quantum Computer can be represented in part as a second-quantized Fermi field.

4. A Quantum Computer in a certain limit naturally forms a Superspace upon which Supersymmetry rotations can be defined – a Continuum Quantum Computer.

5. A representation of Quantum Computers exists that is similar to Turing Machines - a Quantum Turing Machine. As part of this development we define various types of Quantum Grammars.

6. High level Quantum Computer languages are described for the first time.

New linguistic views of the most fundamental theories of Physics, the Standard Model and SuperString Theory are described. In these new linguistic representations particles become literally symbols or letters, and particle interactions become grammar rules. This view is NOT the same as the often-expressed view that Mathematics is the language of Physics. The linguistic representation is a specific new mathematical construct.

We show how to create a SuperString Quantum Computer that naturally provides a framework for SuperStrings in general and heterotic SuperStrings in particular.

There are also a number of new developments relating to Quantum Computers and Quantum Turing Machines that are of interest to Computer Science.

The elementary particle language defines a new form of reality in the sense that the elementary particles become states within a Quantum Computer. The workings and contents of the universe are reduced to computer computation. Speculations about the purposes and goals of the elementary particle language are presented based on the goals and features of languages like C++ and Java.

# 1

# The Standard Model of Elementary Particles

Our understanding of the fundamental nature of matter today is in one way much more developed than that of 1900. In another way we still are at a stage where we think of particles of matter as fuzzy little balls. The balls are different – quarks, leptons, and so on in year 2000 instead of atoms of chemical elements in year 1900. And we know more about their properties. But they are still fuzzy little balls.

The accepted theory of the day is called the Standard Model[1]. It is an amalgam of earlier theories of electromagnetism, the weak interaction and the strong interaction. The Standard Model is consistent with all known experimental results. This combination of existing theories works but it has a number of arbitrary features that cannot be logically justified. It is known to be incomplete because it does not include gravitation.

The Standard Model started with the unification of the theory of Quantum Electrodynamics (electromagnetism) with the theory of the Weak interactions in a theory that is called the Electroweak Theory. This theory was developed by Steven Weinberg, Shelley Glashow, Abdus Salam and others. Afterwards the Electroweak theory was combined with the theory of the Strong interactions to produce the Standard Model.

Up to this point we have been sketchy in our description of the fundamental particles of Nature because there is a widely accepted belief amongst physicists that today's "fundamental particles" may not be fundamental in a deeper theory of matter.

**The Families of Matter**
Today the fundamental particles of nature, as described by the Standard Model, are grouped into three families: fermions, gauge field particles and Higgs field particles. Fermions are particles of half-integer spin that constitute what we ordinarily call matter. The fermions include electrons, quarks, muons, neutrinos and so on.

Gauge field particles are particles that "carry" the interactions or forces between particles. You could call them "force field" particles because they embody the forces between particles. The photon is a gauge field. There are other gauge fields for the weak interaction and for the strong interaction.

Higgs field particles are heavy spin zero particles that are needed in the current formulation of the theory. They have not as yet been observed. The reason is thought to be their high mass. Particle accelerators are only now becoming capable of producing them. There are numerous speculations that Higgs particles are not fundamental but may be some form of composite particle. Since they have not been observed experimentally at the time of this writing there is no way of knowing.

---
[1] There are many books on the Standard Model. See for example Kerson Huang's book, *Quarks, Leptons & Gauge Fields* (World Scientific Publishing, River Edge, NJ, 1992).

The fermion family has two subfamilies: the quarks and the leptons. One distinguishing feature of quarks vs. leptons is that quarks can experience the strong force but leptons cannot. The lepton family includes electrons, muons ($\mu$) and neutrinos ($\nu$). The members of the fermion family are listed in the following table:

**The Fermion Family**

| Generation | Flavor | Quarks | | | | Leptons | |
|---|---|---|---|---|---|---|---|
| I | 1 | up | $u_1$ | $u_2$ | $u_3$ | $\nu_e$ | electron neutrino |
| | 2 | down | $d_1$ | $d_2$ | $d_3$ | e | electron |
| II | 3 | charmed | $c_1$ | $c_2$ | $c_3$ | $\nu_\mu$ | muon neutrino |
| | 4 | strange | $s_1$ | $s_2$ | $s_3$ | $\mu$ | muon |
| III | 5 | top | $t_1$ | $t_2$ | $t_3$ | $\nu_\tau$ | tau neutrino |
| | 6 | bottom | $b_1$ | $b_2$ | $b_3$ | $\tau$ | tau |

Notice also the fermions appear in three generations or sets labeled I, II, and III. There is no obvious reason for this repetition of generations. We do not know at present why Nature has three generations rather than one generation. It is like having three sets of china when one set would do. Each generation contains 6 quarks and two leptons.

Notice that each quark comes in three colors. The colors are labeled with the subscripts 1, 2 and 3. For example the "down" quark actually comes in a triplet – three varieties – that we have denoted $d_1$, $d_2$, and $d_3$ in the above table. These triplets are called color triplets. The variety of quark colors is tied up with the fact that they experience the Strong force.

Originally the three varieties of each quark type were labeled with colors such as red, white and blue. (A "red" quark was not actually red. The name only served to distinguish between the quark varieties.) Now we use simple numeric subscripts. In fact these labels reflect internal quantum numbers that are called color quantum numbers by physicists. Consequently, the theory of the Strong Interactions of the quarks is often called Quantum Chromodynamics since it involves the color quantum numbers.

The sets of color triplets are distinguished from each other by a quantum number called the flavor quantum number. Each triplet of quarks, and a corresponding lepton, has the same flavor quantum number.

The classification of quarks and leptons in the preceding fermion table is the result of over fifty years of experimental and theoretical analysis. It is like the periodic table of the elements that is so important for chemistry.

The Standard Model incorporates all of our current experimental knowledge of particles and interactions (except gravity) in a single theory. The theory has been extremely successful in accounting for many experimental results although many interesting physical calculations in this theory cannot be done with current computational techniques (such as the binding of light quarks to produce protons, neutrons, and so on). In fact, many physicists feel the Standard Model agrees too well with the available experimental data.

The lack of discrepancies and disagreements with experiment is a negative in a sense. Disagreements between experiment and theory are usually the source of advances in our understanding of nature. Grappling with discrepancies leads to new theoretical ideas. Like the ideal marriage with no arguments the success of the Standard Model without experimental discrepancies makes life less interesting for the physicist.

**Current Problems of the Standard Model**

While the Standard Model is not in clear disagreement with any current experimental data there are a number of problems that the Standard Model does not resolve as far as we know. Currently these problems are:

1. Explaining the "dark matter" found in galaxies
   Astronomers have found that the observed galaxies in the universe contain normal matter and an unidentified form of matter that has been given the name "dark matter". Dark matter is undetectable except through its gravitational effects. The amount of dark matter in the universe appears to be about the same as the amount of normal matter. Without dark matter, galaxies would not have enough matter to hold them together and would fly apart because the force of gravity due to visible normal matter is insufficient. Dark matter has the interesting feature that it experiences the gravitational force but it does not experience any of the other forces as far as we know. Light and cosmic rays (elementary particles) appear to pass through dark matter as if it was empty space. Dark matter has some of the interesting properties of the ether of a hundred years ago. A number of schemes have been proposed to account for dark matter. The absence of experimentally verifiable predictions makes it hard to judge whether the correct answer has been found.

2. Incorporating gravity into the Standard Model

The Standard Model is a Quantum Field Theory. Adding the Theory of Gravity as developed by Einstein and others to the Standard Model results in a theory that generates infinitely large numbers (and an infinite number of them at that) when calculations are performed. The normal quantization of the Theory of Gravity as a Quantum Field Theory produces a theory that cannot make finite predictions for perfectly ordinary quantum phenomena and instead produces meaningless infinities. The inability of the Standard Model to incorporate the gravitational interaction without creating difficulties may be a sign of the need for a new theoretical approach.

3. The appearance of "too many" arbitrary constants in the Standard Model theory

The Standard Model contains at least 19 arbitrary constants. These numbers have values that we can determine experimentally but their appearance in the theory is unsatisfying. Physicists hope that there is some deep reason that these numbers have the values found experimentally. This hope leads to the idea that the Standard Model is a precursor to a deeper theory of reality.

1. The lack of "elegance" in the Standard Model

The Standard Model has been described as three theories lumped together: Quantum Electrodynamics, the Theory of the Weak Interaction, and the Theory of the Strong Interaction (called Quantum Chromodynamics). Although the Electroweak (Quantum Electrodynamics combined with the Weak Interactions) part of the Standard Model has some intertwined structure, Quantum Chromodynamics is simply "tossed into the pot" to create the combined theory. An elegant theory would be a closely interconnected theory with all the parts intertwined in a significant way. The close interconnections are a missing ingredient that fills physicists with misgivings about the Standard Model and leads them to view the Standard Model as an interim theory.

2. The "weird" pattern of symmetries in the Standard Model

The Standard Model contains a set of symmetries SU(3) for Quantum Chromodynamics, and SU(2) and U(1) for the Electroweak part that are not easy to justify. These symmetries U(1), SU(2), and SU(3) are symmetry groups chosen from a vast (infinite) number of possibilities appearing in a branch of Mathematics called Lie Groups after their originator Sophus Lie. Why Nature chose these groups is completely unclear. Ideally the "correct" theory would contain one symmetry group that we could somehow justify on physical grounds. This "correct" theory would then, through some mechanism, split into these separate symmetries in the approximate theory called the Standard Model. SuperString Theory is one attempt to create such a theory.

Despite these problems the Standard Model is an extremely successful theory and a triumph for the physicists – experimental and theoretical – whose labors over the last seventy years led to its creation.

In subsequent sections of this chapter we will look at some of the interactions between particles in the Standard Model. The interactions involve the creation and absorption of particles. We will also look at these interactions from a linguistic point of view in keeping with the general theme of this book that Theoretical Physics is in reality a search for a language with which to describe Nature. The approach followed here will tie in with the discussions of quantum computers in succeeding chapters.

**Interactions of the Standard Model**

The Standard Model consists of a complex mathematical expression called a *lagrangian* together with a procedure for quantization, and a procedure for the calculation of physically interesting quantities. The quantization and calculation procedures are specified by Quantum Field Theory. The details of these procedures are important but they are not relevant to the theme of this book.

The lagrangian for the Standard Model can be divided into two parts. One part describes the behavior of non-interacting particles. These particles are called *free particles*. After some mathematical development we can think of free particles as moving through space and interacting with other free particles through the exchange of gauge particles. Gauge particles are the particles that are the carriers of the electromagnetic, weak, and strong forces. The photon is the gauge particle for Quantum Electrodynamics (electromagnetism). Three W bosons are the gauge particles for the Weak Interaction. (The photon and the W particles are intertwined in the Electroweak theory.) Eight gluons are the gauge particles for Quantum Chromodynamics (the Strong Interaction).

The second part of the lagrangian describes the interactions of particles. These interactions typically involve the exchange of gauge particles between the interacting particles. The interactions on the quantum scale become the forces we see in everyday life when vast numbers of quantum interactions take place.

> It is an interesting point of history to note that hundreds of years ago scientists thought that light was made of particles and that seeing involved the reception of particles. Forces were also viewed as the result of myriads of particles acting on a body. These ideas were discarded in the nineteenth century only to be revived in the latter part of the twentieth century in a much deeper and more detailed quantitative form.

The interaction part of the Standard Model specifies many interactions between particles. We will look at a representative sample of these interactions to obtain an understanding of their basic idea and then use these interactions to illustrate the linguistic representation of particle interactions. The linguistic approach will be explored in more detail in succeeding chapters.

The first interaction term in the lagrangian that we will explore can be written in the form:

$$\bar{e}Ae$$

where e represents an electron (the bar over the left e signifies an "outgoing" electron in our simplified view) and A represents a photon. (Many significant features are omitted to avoid getting bogged down in details that are not central to the theme of this book.) This term represents a number of different interaction situations. The situations can be depicted graphically as:

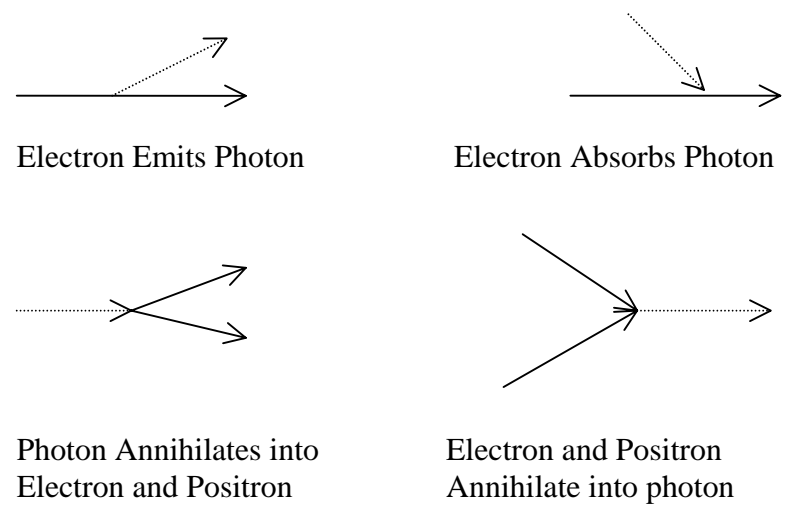

Electron Emits Photon                Electron Absorbs Photon

Photon Annihilates into              Electron and Positron
Electron and Positron                Annihilate into photon

Figure. Simplest Electromagnetic Interactions of an electron. Solid lines are electrons and positrons (anti-electrons). Dashed lines are photons.

The interactions depicted in the figure correspond to an amazing feature of matter: particle creation and annihilation with the conversion of energy into matter and vice versa. This energy-matter conversion requires both the quantum theory and the theory of Special Relativity to work. Special Relativity is required because of the conversion between matter and energy. The creation and annihilation process is an inherently quantum process requiring Quantum Field theory.

The interaction processes in the preceding figure are the simplest electromagnetic interactions of an electron. These simple interactions can be combined to make an infinity of more complex composite interactions. For example two electrons can scatter off each other by exchanging a photon. The photon exchange takes place by having one electron emit a photon and the other electron absorb it. This composite process can be pictured with a (Feynman) diagram such as:

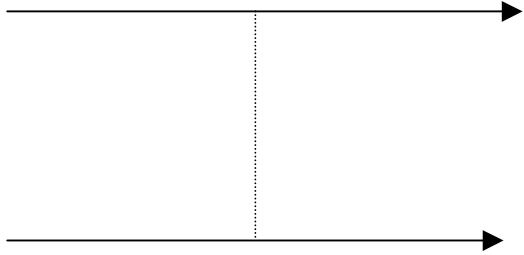

Figure. Feynman diagram for two electrons interacting by exchanging a photon (dotted line).

The preceding discussion shows how the simple interactions in the Standard Model lagrangian can produce a complex variety of phenomena. The Feynman diagrams that we have seen are used extensively in particle physics. They are named after their originator Richard Feynman. These diagrams correspond to very complex rules for performing calculations in quantum field theory. The calculations produce probability numbers giving the probability that a given process will take place.

**Perturbation Theory and a Language for the Standard Model**

This brief view of some of the features of the Standard Model is a preliminary to a new view or representation of the physics of the Standard Model. *In this view the interaction terms in the lagrangian specify the grammar of a computer language.*

Physicists must use an approximation procedure called perturbation theory to perform computations in the Standard Model. Perturbation theory uses the interaction terms of the lagrangian to perform an approximate calculation of particle interactions. Perturbation theory calculations are normally visualized using Feynman diagrams. For example the collision of two electrons to produce two electrons with different energy and momenta can be visualized as a sum of terms corresponding to the various ways the electrons can interact. This sum of terms is an approximate perturbation theory calculation. It is approximate because the exact answer would require calculating an infinite number of terms – and nobody knows how to perform such a calculation. The simplest and dominant terms in electron-electron scattering are:

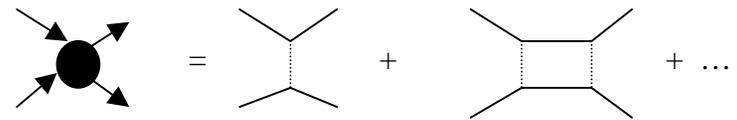

Figure. The first few terms in the approximate perturbation theory calculation of the scattering of two electrons. The dotted lines represent photons that carry the electromagnetic force between the electrons.

The individual terms in the perturbation theory calculation can be viewed as words. The words are part of a language with an alphabet and grammar.

The 36+ particles of the Standard Model constitute the set of symbols or the alphabet of the language. (It is an interesting coincidence that most alphabet-based human languages have 20 to 40 letters in their alphabets. English has 26 and so on.) As we shall see, the grammar is quantum extension of a type of computer language (developed by Chomsky and others) that uses *production rules*. The production rules for the grammar of the Standard Model are easily derived from the interaction terms of the Standard Model.

The concept of the language of the Standard Model is very simple. The technical details of the language description require a discussion of computer languages and Quantum Turing Machines. The next few chapters describe the basic idea of the linguistic representation of the Standard Model. They show how particle interactions can be viewed as transformations (processes) within a Quantum Computer that accepts the computer language generated by the Standard Model lagrangian interaction terms.

# 2

# The Linguistic Representation of the Standard Model

In the previous chapter we explored the features of the Standard Model. We saw that it was consistent with all the known properties of elementary particles except for gravitation, and phenomena such as dark matter. This chapter introduces a new view or representation of the Standard model that focuses on its interactions and *shows the Standard Model defines a language similar to a computer language*.

The alphabet of the language is the set of elementary particles of the Standard Model. The words of the language are quantum states consisting of elementary particles. A bound state of several particles such as a proton (three quarks bound together) is a word. A quantum state consisting of several free particles – particles that are not bound together and that are some distance from each other also constitute a word. In fact the entire universe constitutes one mighty word.

We shall see that the collision or scattering of particles can be viewed as beginning with a combination of letters corresponding to the particles to make an input string. This input string undergoes transformations specified by grammar rules to produce an output string of letters corresponding to the outgoing particles after the collision.

In describing this new representation of the Standard Model we will focus on the essentials of the processes of creation, transmutation and annihilation of matter ignoring (for the moment) particle spin, momentum, angular momentum and other details that are important in the complete theory of the Standard Model. (These details will be considered when we consider the SuperString Quantum Computer. Incorporating them into a language is not difficult.)

The idea of associating physics with computers is not as unconventional as it might appear at first. Feynman[2] viewed computers as relevant for Physics: "If we suppose we know all the physical laws perfectly, of course we don't have to pay any attention to computers. It's interesting anyway to entertain oneself with the idea that we've got something to learn about physical laws; and if I take a relaxed view here … I'll admit that we don't understand everything." Feynman wanted to simulate physics computations on a quantum computer in the hope that it would be faster than a conventional computer. We will show the Standard Model itself actually defines a specific (theoretical) quantum computer – a far more exciting possibility – because it gives a new view of Reality. Nature itself is a form of computer. We will also show that SuperString theory can be formulated within a Quantum Computer framework.

A computer language representation of particle physics is of great interest in itself. It may generate new insights into the process of matter creation and transformation. It may lead to a new understanding of the fundamental nature of the universe. And it appears to suggest a rationale for approaches such as the currently popular SuperString theories of elementary particles.

---

[2] R. P. Feynman, International Journal of Theoretical Physics, **21**, 467 (1982).

## Linguistic View of an Interaction

We will begin by looking at the simple interaction term we explored in the previous chapter:

$$\bar{e}Ae$$

From a computer language perspective this lagrangian interaction term can be viewed as specifying a set of grammar rules called *production rules*.

In fact each interaction term in the Standard Model lagrangian can be viewed as specifying a set of grammar rules. The combined set of grammar rules for all Standard Model interaction terms constitutes a grammar equivalent to the Standard Model lagrangian interaction terms with particles becoming the alphabet (letters or symbols) of the grammar.

To appreciate the mapping or analogy of particles and alphabetic letters, and of interaction terms and computer grammar, we have to understand the process of data characters (or letters) flowing through a computer. It is an interesting and little noted fact (because it is viewed as trivial) that a computer can generate (or absorb) data as part of the computation process. For example we might write a computer program that takes a set of letters input into a computer and outputs each input letter twice:

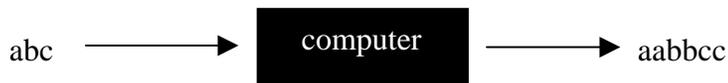

*In a sense the computer creates data characters just like interactions can create particles.* Computers can also absorb (or annihilate) data (usually to our dismay). So we can see an analogy between the transformations of data characters in a computer, and particle annihilation and creation. *Nothing else in Nature is so directly analogous to particle creation and annihilation.* We take the view that particles are bundles of data that we can denote with letters (symbols). They contain quantum numbers and other properties (mass, spin, and so on) that certainly are data.

## Computer Grammars

The Standard Model lagrangian in our view specifies a *grammar* in the sense of Naom Chomsky. Chomsky's concept of a language, and of a grammar, has important applications in the theory of computation and computers.

There are four basic types of languages in the Chomsky approach: type 0, type 1, type 2 and type 3. They differ in the forms of the grammar rules (also called *production rules*) that they allow. We will be interested in type 0 languages. A type 0 language (also called an *unrestricted rewriting system*) is the most general type of language. It allows any grammar production rule of the form

$$x \longrightarrow y$$

where x and y are strings of characters.

Production rules specify how one string of characters transforms into another string of characters. Calculations in computers using computer languages are reducible to sets of grammar rules for string manipulation that are similar to the one shown above.

Each term in the interaction part of the Standard Model lagrangian is equivalent to one or more production rules where the characters are particles. *The Standard Model can be viewed as generating a type 0 language.* This language goes beyond current types of grammars because it is inherently quantum probabilistic in nature. Quantum aspects of these rules will be described later.

Before looking at the production rules generated by an interaction term in a lagrangian we will look at a formal grammar. A grammar is a quadruple of items that is symbolized by the expression

$$\texttt{<N, T, S, P>}$$

where N is a set of variables called *nonterminal symbols*, T is a set of *terminal symbols*, S is a special nonterminal symbol called the *head* or *start symbol*, and P is a finite set of production rules. The angle braces **<** and **>** are merely a mathematician's way of saying these items are grouped together to constitute (or make) a grammar.

The terminal symbols are the set of characters that are allowed in input strings or output strings. The nonterminal symbols are the set of characters that appear in intermediate steps that lead from the input string to the output string. They are like internal variables or symbols. The combined set of terminal and nonterminal symbols make up the *vocabulary* (alphabet) of a language.

Chomsky's definition of a language is the set of all strings of terminal symbols that can be generated by applying the production rules to the head symbol (or start symbol) S. The head symbol is the symbol that begins all strings of symbols that can be generated in the language.

A simple example of a language in this approach is a vocabulary or alphabet consisting of the ABC's with words created from these letters according to some set of production rules.

We will generalize Chomsky's idea of language to be the set of all strings that can be generated from all input strings of terminal symbols as well as the *head symbol*. We can also view all particles as generated directly or indirectly at the beginning on the universe. The "Big Bang" (the beginning of the universe) then becomes the primeval head symbol.

A visualization of the application of production rules to transform an input string of terminal characters into an output string of terminal characters is:

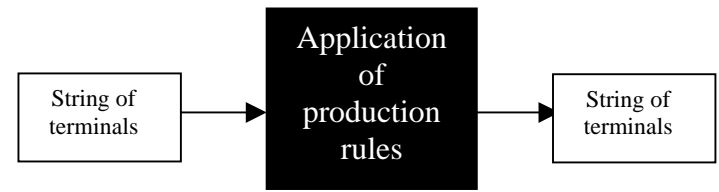

Figure. Generating an output string of terminal characters from an input string of terminal characters using the production rules of a grammar. Inside the black box transformations of the input string take place and non-terminal symbols may appear and disappear. Non-terminal symbols, by definition, cannot appear in the input or output strings of characters.

In order to make these grammar concepts more concrete we will look at a simple artificial grammar before looking at the grammar generated by interaction terms in the Standard Model. The nonterminal symbols will be the letters S (the head symbol), A and B. The terminal symbols will be the letters x and y. The production rules will be

$$S \to AB \quad \text{Rule I}$$
$$A \to y \quad \text{Rule II}$$
$$A \to Ay \quad \text{Rule III}$$
$$B \to x \quad \text{Rule IV}$$
$$B \to Bx \quad \text{Rule V}$$

The computer language that this grammar generates consists of all strings containing any number of y's followed by any number of x's since any of these strings can be generated from the head symbol S using the production rules. Because of rule I y's are placed to the left and x's are placed to the right. The order of the symbols matters just as it does in human language – consider the words ides and dies which differ only in the order of i and d!

An example of generating a string yyxxx from the head symbol is:

$$\begin{aligned}
&S \rightarrow AB && \text{Rule I} \\
&AB \rightarrow AyB && \text{Rule III} \\
&AyB \rightarrow yyB && \text{Rule II} \\
&yyB \rightarrow yyBx && \text{Rule V} \\
&yyBx \rightarrow yyBxx && \text{Rule V} \\
&yyBxx \rightarrow yyxxx && \text{Rule IV}
\end{aligned}$$

The production rule used to make each transition is listed on the right.

Our generalization of the Chomsky definition of language would allow any string to be the starting point – not just the head symbol S. Using the sample grammar described on the previous page the generalized language becomes any string of x's and y's. To make it more interesting we can add two new rules:

$$\begin{aligned}
&y \rightarrow A && \text{Rule VI} \\
&x \rightarrow B && \text{Rule VII}
\end{aligned}$$

The language – the set of strings of terminal symbols – remains the same despite the addition of these new grammar rules. However the number and variety of transitions becomes much larger. For example the following chain of transitions is allowed,

$$yx \rightarrow AB \rightarrow AyBx \rightarrow AyyBx \rightarrow AyyyBxx \rightarrow yyyyxx$$

# 3

# Probabilistic Computer Grammars

**Probabilistic Computer Grammars**
The preceding chapter described the production rules for a deterministic grammar. The left side of each production rule has one, and only one, possible transition.

Non-deterministic grammars allow two or more grammar rules to have the same left side and different right sides. For example,

$$A \to y$$
$$A \to x$$

could both appear in a non-deterministic grammar.

Non-deterministic grammars are naturally associated with probabilities. The probabilities can be classical probabilities or quantum probabilities. An example of a simple non-deterministic grammar is:

The head symbol is the letter S. The terminal symbols are the letters x and y. The production rules are:

$S \to xy$      Rule I
$x \to xx$      Rule II   Relative Probability = .75
$x \to xy$      Rule III   Relative Probability = .25
$y \to yy$      Rule IV

The probability of generating the string xxy vs. the probability of generating the string xyy from the string xy is

$xy \to xxy$      relative probability = .75

<p style="text-align:center">vs.</p>

$xy \to xyy$      relative probability = .25

The string xxy is three times more likely to be produced than the string xyy.

For each starting string one can obtain the relative probabilities that various possible output strings will be produced.

A more practical example of a Probabilistic Grammar can be abstracted from flipping coins – heads or tails occur with equal probability – 50-50. From this observation we can create a little Probabilistic Grammar for the case of flipping two coins. Let us let h represent heads and t represent tails. Then let us choose the grammar:

$S \to hh$
$S \to tt$
$S \to ht$
$S \to th$
$h \to t$      probability = .5 (50%)
$h \to h$      probability = .5 (50%)
$t \to h$      probability = .5 (50%)

t → t          probability = .5 (50%)

The last four rules above embody the statement that flipping a coin yields heads or tails with equal probability (50% or .5).

Now let us consider starting with two heads hh. The possible outcomes and their probabilities are:

hh → hh      probability = .5 * .5 = .25
hh → th      probability = .5 * .5 = .25
hh → ht      probability = .5 * .5 = .25
hh → tt       probability = .5 * .5 = .25

If we don't care about the order of the output heads and tails then the probability of two heads hh → ht or th is .25 + .25 = .5.

This simple example shows the basic thought process of a non-deterministic grammar with associated probabilities.

The combination of a non-deterministic grammar and an associated set of probabilities for transitions can be called a *Probabilistic Grammar*. We will see that the grammar production rules for the Standard Model must be viewed as constituting a Probabilistic Grammar™ with one difference. The "square roots" of probabilities – probability amplitudes are specified for the transitions in the grammar. Probability amplitudes are required by the Standard Model because it is a quantum theory. Therefore we will call the grammar of the Standard Model a *Quantum Probabilistic Grammar™*.

**Quantum Probabilistic Grammar**
An example of a Quantum Probabilistic Grammar™ can be constructed in analogy with a two slit photon experiment. Imagine a wall with two slits. A source of photons shoots photons at the wall. A photon can go though either slit with equal quantum probability. An illustration of this experimental arrangement is:

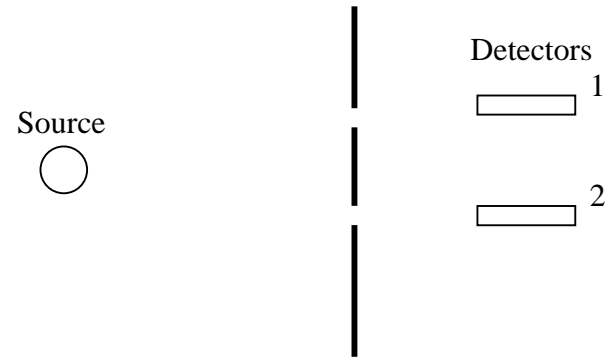

Figure. Two slit photon experimental setup.

A simple Quantum Probabilistic Grammar can be constructed corresponding to this experimental setup:

S → 1    probability amplitude = $1/\sqrt{2}$
S → 2    probability amplitude = $1/\sqrt{2}$

The head symbol S represents the source. The digit 1 represents a photon going through slit 1. The digit 2 represents a photon going through slit 2.

The values of the probability amplitudes $1/\sqrt{2}$ can be calculated using Quantum Mechanics. The probability for a photon to go through slit 1 is the absolute value squared of the probability amplitude:

Probability to go through slit 1 = $(1/\sqrt{2})^2 = .5$

and the probability for a photon to go through slit 2 is

Probability to go through slit 2 = $(1/\sqrt{2})^2 = .5$

This simple example illustrates the basics of a Quantum Probabilistic Grammar™.

Before applying these concepts to the Standard Model we will look at a simpler Quantum Field Theory called a $\phi^3$ ("phi cubed") theory ($\phi$ is the Greek letter phi). This theory describes a self-interacting spin 0 particle with no internal symmetries. This artificial theory is a stepping stone to the far more complex Standard Model Quantum Field Theory.

The $\phi^3$ theory is so named because it has a lagrangian interaction term with that form. The grammar rules for the $\phi^3$ theory are:

$$\phi \to \phi\phi \quad \text{Rule I}$$
$$\phi\phi \to \phi \quad \text{Rule II}$$

Rule I corresponds to the emission of a $\phi$ particle and rule II corresponds to the absorption of a $\phi$ particle. We do not introduce a start symbol. Instead we will consider the transitions from an input state of a number of $\phi$ particles to an output state of possibly a different number of $\phi$ particles. We will ignore the momenta of the particles. (This assumption is equivalent to infinitely massive particles.) We will assume either transition above takes place with a "relative probability amplitude" g. We will call this simplified theory the *modified $\phi^3$ theory*. We will view g as a measure of the probability amplitude for an absorption or emission of a $\phi$. (g is similar to a coupling constant in Quantum Field Theory.) The actual summed probability has to be normalized or rescaled so that the sum of probabilities equals one.

To get a feel for the Quantum Probabilistic Grammar™ approach we will look at the case of an input state of two $\phi$ particles. The output states can have one $\phi$, two $\phi$'s, three $\phi$'s, and so on. Each possible output state has a certain probability of occurring. The sum of the probabilities of producing all output states must equal one. (Remember that the sum of all possible outcomes of flipping a coin is one. Having it come up heads has probability ½ and having it come up as a tails has probability ½ also. And ½ plus ½ equals one.)

The simplest string transition from a two $\phi$ "input" state to a one $\phi$ "output" state is:

$$\phi\phi \to \phi$$

using Rule II. The probability amplitude of this transition is g by assumption.

The transitions between strings can be visualized with diagrams that are like the Feynman diagrams that used in Quantum Field Theory perturbation theory calculations. These diagrams are not the same as Feynman diagrams because they embody time orderings of emissions and absorptions of $\phi$ particles.

The time order of emission and absorption of $\phi$ particles can be symbolized using parentheses. For example,

$$(\phi)\phi \to (\phi\phi)\phi = \phi(\phi\phi) \to \phi(\phi) = \phi\phi \qquad \text{Diagram A}$$

and

$$\phi(\phi) \to \phi(\phi\phi) = (\phi\phi)\phi \to (\phi)\phi = \phi\phi \qquad \text{Diagram B}$$

These string transitions correspond to different time-ordered Feynman-like diagrams:

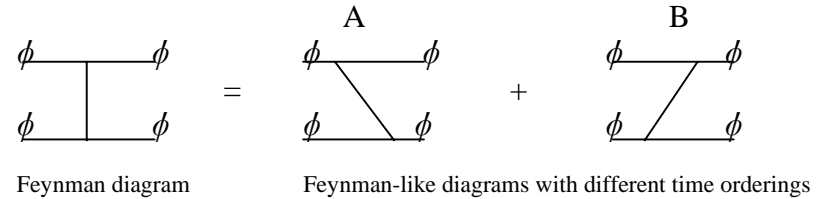

Feynman diagram        Feynman-like diagrams with different time orderings

Figure. The true Feynman diagram is the "sum" of the two time-ordered Feynman-like diagrams.

The correspondence between the Feynman-like diagrams and the transitions between strings based on the Quantum Probabilistic

Grammar™ can be seen by taking vertical slices on diagrams A or B above after each emission or absorption. For example,

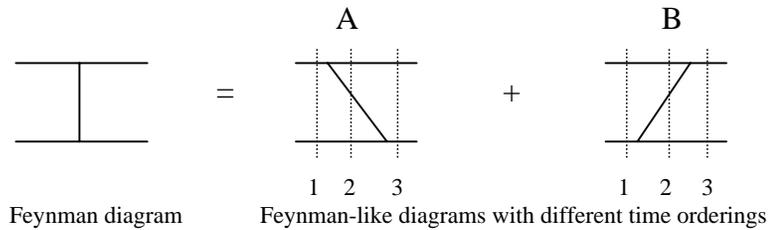

Feynman diagram     Feynman-like diagrams with different time orderings

Figure. Slicing Feynman-like diagrams. A slice is made after each emission or absorption. As you read down a slice the particles are listed in the same order as the corresponding string. Each slice is numbered.

The string corresponding to each slice in the above figure is numbered in the transitions:

$$\overset{1}{(\phi)\phi} \to \overset{2}{(\phi\phi)\phi} = \phi(\phi\phi) \to \overset{3}{\phi(\phi)} = \phi\phi \qquad A$$

and

$$\overset{1}{\phi(\phi)} \to \overset{2}{\phi(\phi\phi)} = (\phi\phi)\phi \to \overset{3}{(\phi)\phi} = \phi\phi \qquad B$$

The parentheses on the left side of the arrow indicate the particle that emits the new particle appearing within the parentheses on the right side of the arrow.

A transition from an input state containing $\phi$ particles to an output state containing $\phi$ particles always has an infinite number of ways of taking place and thus an infinite number of Feynman-like diagrams. Readers familiar with perturbation theory in Quantum Field Theory can see that these diagrams are the same as the Feynman diagrams generated by perturbation theory with the additional feature of having time orderings.

We will now look at the transition of two $\phi$ particles to two $\phi$ particles: $\phi\phi \to \phi\phi$. There are an infinite number of Feynman-like diagrams for this transition.

Now in evaluating diagrams to calculate the probabilities we must remember that we are ignoring space-time aspects such as particle propagators. So the calculation of the probability amplitude for this process becomes a counting of the number of diagrams that exist for each power of $g^2$. The probability amplitude for each diagram is a power of g.

Counting diagrams is a combinatorial mathematics problem that we will not explore in detail because it is peripheral to our interests. Consequently we will simply write the probability amplitude as:

$$A_2(g) = \sum_{n=1}^{\infty} a_n g^{2n}$$

where the mathematical expression on the right represents a sum from n = 1 to infinity and where the numbers $a_n$ are integer numbers equal to the number of different diagrams having a power of $g^{2n}$. Each intersection of lines (called a vertex) contributes a factor of g to the amplitude for that diagram. The powers of g for the simpler diagrams are:

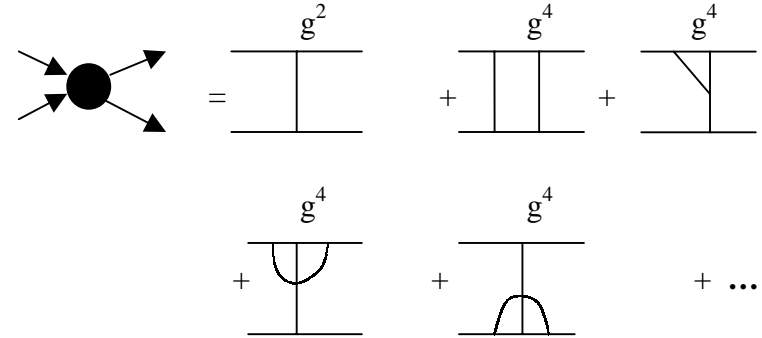

The value of the first constant $a_1$ is 1 since there is only one Feynman diagram with value $g^2$ for this process – the first diagram to the right of

the = sign above. We will treat all time-ordered variations of a Feynman diagram as contributing one to the value of $a_n$. The value of $a_n$ grows rapidly as n increases. For large n the value of $a_n$ is of the order of $(n!)^2$. Consequently the sum is an asymptotic power series. Here again the details are not important. We are not looking for a numerical result.

The (unnormalized) relative probability for the transition $\phi\phi \to \phi\phi$ is:

$$P_2 = |A_2(g)|^2$$

In quantum theories the probability is always the square of the absolute value of a probability amplitude. The probability $P_2$ is a relative probability that must be normalized – multiplied by a factor that makes the sum of the probabilities of all possible outcome states equal to one. To calculate this probability we must calculate the sum of all the relative probabilities $P_n$ to produce any number of $\phi$ particles from a two $\phi$ input state.

$$\phi\phi \to \phi \ldots$$

The total of the relative probabilities is:

$$P = \sum_{n=1}^{\infty} P_n$$

where $P_n$ is the relative probability to produce an output state with n $\phi$ particles.

The calculation of the relative probabilities $P_n$ for n $\phi$ particles output states is similar to the calculation $P_2$. For example,

$$A_3(g) = \sum_{n=1}^{\infty} b_n g^{2n+1}$$

where the numbers $b_n$ count the number of distinct diagrams with the power $g^{2n+1}$ and

$$P_3 = |A_3(g)|^2$$

The absolute (normalized) probability to produce an n $\phi$ particle output state is

$$Q_n = P_n/P$$

The sum of all possible output state probabilities equals one:

$$1 = \sum_{n=1}^{\infty} Q_n$$

The modified $\phi^3$ Quantum Field Theory provides a simple example of a Quantum Probabilistic Grammar™. We will now turn to the Standard Model and examine its Quantum Probabilistic Grammar™. Because it encompasses a much larger number of different particles (letters) and interactions (grammar rules) it will be significantly more complicated.

# 4

# Standard Model Quantum Grammar™

**Grammar Production Rules of Electromagnetism**
We will start by considering electromagnetic interactions in the Standard Model as grammar production rules. The production rules for the electromagnetic interaction term for electrons and positrons in the Standard Model

$$\bar{e}Ae$$

are:

$$e \to eA$$
$$e \to Ae$$
$$eA \to e$$
$$Ae \to e$$
$$p \to pA$$
$$p \to Ap$$
$$Ap \to p$$
$$pA \to p$$
$$ep \to A$$
$$pe \to A$$
$$A \to ep$$
$$A \to pe$$

where e represents an electron, p represents a positron, and A represents a photon. The production rules describe the emission and absorption of photons by electrons and positrons as well as the annihilation of an electron and positron to produce a photon, and the decay of a photon into an electron-positron pair.

An example of an interaction between two electrons in the linguistic approach is:

$$\overset{1}{ee} \to \overset{2}{eAe} \to \overset{3}{ee}$$

where the electrons interact by exchanging one photon. One Feynman-like diagram for these transitions is:

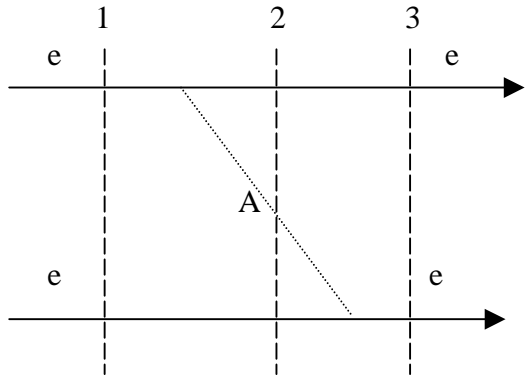

Figure. A diagram showing how two electrons interact by exchanging a photon. As time increases the electrons move from left to right. The upper electron emits the photon. This corresponds to the left e transitioning to eA using the grammar rule e → eA.

The vertical slices in the Feynman diagram which are numbered 1, 2 and 3 correspond to the three numbered strings in the transitions generated from the production rules. Each string has an ordering that corresponds to the order of particles as you descend a slice. For example slice 2 has an electron, photon, and another electron in that order as you descend corresponding to string 2 above.

Another Feynman-like diagram that is important and contributes to this process has the lower electron emitting a photon that is then absorbed by the upper electron.

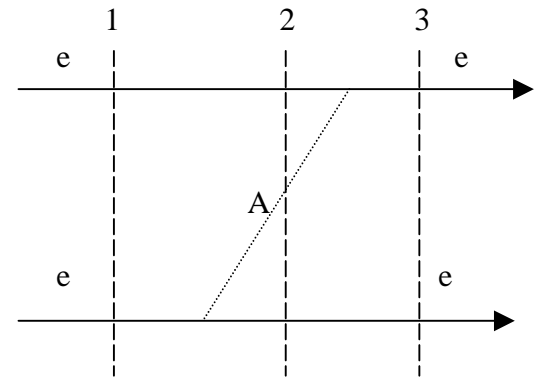

Figure. Another diagram with two electrons interacting by exchanging a photon. In this case the lower electron emits the photon. This corresponds to the right e in the initial ee string transitioning to Ae using the grammar rule e → Ae.

The "normal" Feynman diagram for this process represents the sum of both of the previous diagrams:

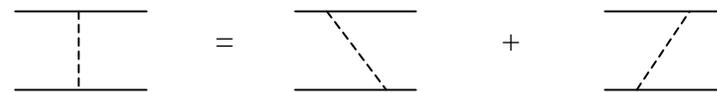

Feynman diagram      Feynman-like diagrams with different time orderings

Figure. The normal Feynman diagram represents several of our Feynman-like diagrams with different time orderings of particle emission and absorption.

A more complex example of a Feynman-like diagram is:

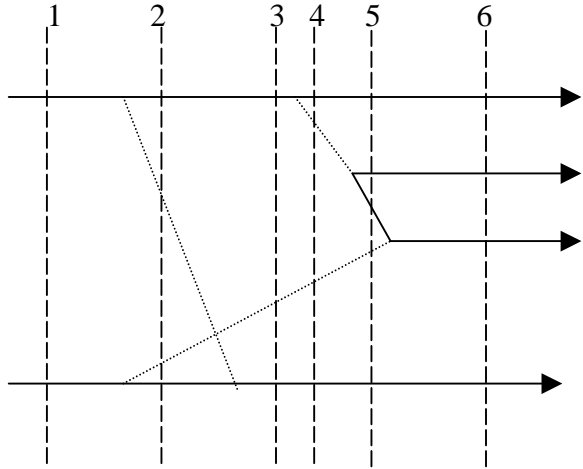

Figure. A diagram for the collision of two electrons that produce a new electron-positron pair: ee → eepe.

Six slices appear corresponding to various intermediate states in this complex electron-electron interaction. The production rules can be used to generate a sequence of strings that correspond to the slices:

    1      2      3      4      5      6
  ee → eAAe → eAe → eAAe → eepAe → eepe

As you descend each slice the particles are ordered in the same way as the corresponding string. For example, as you descend slice 5 the order of the particles is electron, electron, positron, photon and electron.

The transitions of character strings have an ambiguity. For example the above transition

          2      3
     eAAe → eAe

could have taken place through (eA)Ae → (e)Ae with the left e absorbing an A or through eA(Ae) → eA(e) with the right e absorbing an A. (Parentheses are used for grouping to show which electron absorbed the photon.) This ambiguity reflects the fact that there are several possible time orderings. The preceding diagram actually corresponds to eA(Ae) → eA(e). The right electron absorbs the photon.

To calculate the probability of an actual physical transition taking place such as ee → eepe we must take account of all diagrams with all possible time orderings for the specified input and output states.

The above examples show how the electromagnetic interaction part of the Standard Model lagrangian can be viewed as defining a grammar. The grammar has corresponding Feynman-like diagrams. These types of diagrams are not without precedent in Physics. They have time orderings that are similar to the time orderings that appeared in perturbation theory calculations around 1950.

The Weak Interaction and Strong Interaction parts of the Standard Models also define grammars. Consequently we can view the complete Standard Model lagrangian as defining a grammar where the "letters" (alphabet or vocabulary) are the elementary particles of the model and the Feynman-like diagrams corresponding to the Standard Model can be viewed as a sequence of strings generated by applying the production rules specified by the Standard Model lagrangian.

**Production Rules for the Weak and Strong Interactions**

The Weak and Strong interaction terms in the Standard Model lagrangian are also easily translated into grammar production rules. We will illustrate these cases using the Weak interaction terms:

$$\bar{\nu}_e W^- e$$

and

$$\bar{\nu}_\mu W^- \mu$$

where $\nu_e$ represents an electron neutrino, $\nu_\mu$ represents an muon neutrino, $\mu$ represents a muon, $W^-$ is a gauge field of the Weak interaction and e is an electron; and the Strong interaction term

$$\overline{u}Gu$$

where u is a u quark and G represents gauge fields of the Strong interaction.

Notice that neutrinos come in different varieties: electron neutrinos, muon neutrinos and tau neutrinos. The three kinds of neutrinos have different internal quantum numbers that distinguish them. Neutrinos do not have electromagnetic charge. They are neutral as their name suggests. Each kind of neutrino has a corresponding charged partner. We are familiar with the electron. The other charged partners are the muon and tau particle. These particles are like heavy electrons for the most part. The three charged leptons have different internal quantum numbers.

The above interaction terms imply production rules such as:

$$e \rightarrow W^- \nu_e$$
$$e \rightarrow \nu_e W^-$$
$$W^- \rightarrow e\, \nu_e$$
$$W^- \rightarrow \nu_e\, e$$
$$\mu \rightarrow \nu_\mu W^-$$
$$\mu \rightarrow W^- \nu_\mu$$
$$u \rightarrow G u'$$
$$u \rightarrow u' G$$

and so on where e is an electron, p is a positron, $W^-$ is a negative W gauge boson, $\nu$ is a neutrino, G is a Strong interaction gauge boson and u and u' are u quarks which may have different color quantum numbers.

These production rules generate Weak interaction transitions such as muon decay:

$$\overset{1}{\mu} \rightarrow \overset{2}{W^- \nu_\mu} \rightarrow \overset{3}{e \nu_e \nu_\mu}$$

with the corresponding Feynman-like diagram:

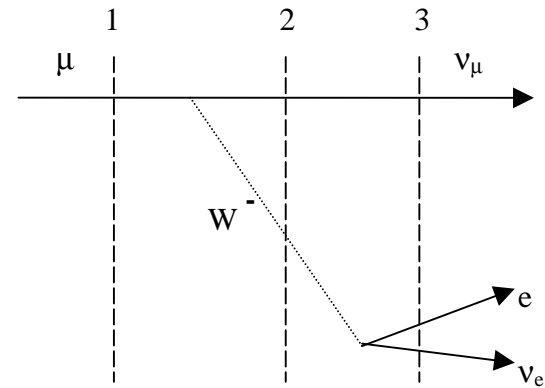

They also generate Strong interactions between quarks with color gauge fields:

$$\overset{1}{uu} \rightarrow \overset{2}{uGu} \rightarrow \overset{3}{u'u'}$$

with the corresponding Feynman-like diagram:

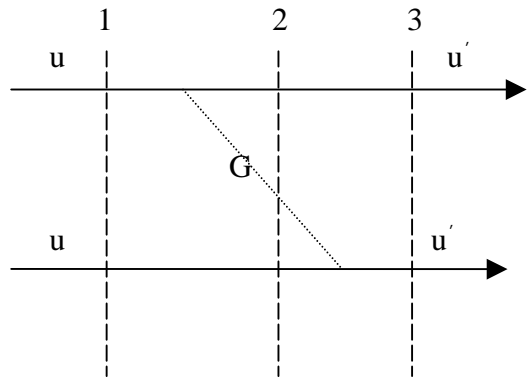

The preceding examples are the simplest cases of the infinite variety of Feynman-like diagrams that can be generated from the Standard Model production rules.

**The Standard Model Quantum Grammar**
At this point we have created a view of the Standard Model of elementary particles in which the particles are an alphabet of at least thirty-six letters (eighteen quarks, six leptons, twelve gauge fields, plus possibly other particles such as Higgs bosons).

The particle alphabet can be combined into strings that represent input or output states of scattering particles. Transitions between strings take place through the grammar production rules and correspond to time-ordered Feynman-like diagrams.

So we now have a language, the Standard Model, with letters, words and a grammar, and an interpretation of the language in terms of rules of calculation and experimental setups.

The linguistic representation of the Standard Model that we have developed omits many important calculation details as well as important properties of particles such as spin and particle momenta. These features could be added in a direct way. The focus of our investigation is on the essentials of the act of creation of particles in particle interactions.

The character string transition approach based on production rules is equivalent to the Feynman diagram approach. It does however provide a different, and simpler, view. Interestingly the Feynman diagram approach to calculations is very difficult and tedious – but it often leads to a simple result due to the massive cancellation of many complex terms with each other. (An attempt to simplify perturbation theory diagram calculations was made by Wu and Cheng in the early 1970's (and by this author privately). The simple linguistic approach might be a hint of a more efficient way of calculating in field theories like the Standard Model where the complications are absent from the very beginning.

Whether or not the linguistic approach leads to a less complicated theory we have now reached a rather amazing result. After 2500 years of speculation on the nature of matter we have developed a surprisingly simple theory (for everything except gravitation) called the Standard Model that can be viewed as a type 0 computer language. It has an alphabet (vocabulary) of roughly 36 or so particles, and a set of production rules specified by the interaction part of the Standard Model lagrangian.

This situation represents something of a miracle. There is no reason that Nature should have so few particles that interact with each other through a simple set of rules. A computer language theorist would call the language of the Standard Model a language with a finite representation. Simply put, this means the words of the language can be generated from a finite vocabulary (alphabet or set of particles) and a finite set of production rules.

**The Standard Model Language is Surprisingly Simple**
Many physicists feel that there are too many elementary particles in the Standard Model. From the point of view of a language theorist *the finite language representation of the Standard Model is a very special situation*. As Hopcroft and Ullman[3] point out, "there are many more languages than finite representations." Languages can have infinite alphabets or infinite sets of production rules or other complications.

---
[3] J. E. Hopcroft and J. D. Ullman, *Formal Languages and Their Relation to Automata*, (Addison-Wesley, Reading, MA, 1975) page 2.

*The physical equivalent of an infinite alphabet would be a universe with an infinite number of different types of matter. Every particle of matter could have a different mass and differ in other properties. From this point of view the Standard Model is truly a marvel of simplicity.*

The Standard Model, in fact, is a very compact, finite description of most of the known features of our universe. The linguistic view of the Standard Model suggests we should view elementary particles as symbols or clumps of data – a vocabulary. The interactions of the elementary particles involve the creation or annihilation of particles – creation in the deepest form seen by man.

Remarkably, the data flowing through a computer can be viewed as being transformed from one form to another and being output to different destinations. Data flowing through a computer can be divided into streams that can be sent to different output channels such as a printer or the computer screen. For example, the character, a, can be a data item in a stream of data that is sent to both the printer and the screen:

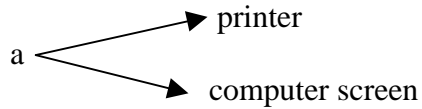

The streaming of data characters to different output channels is very analogous to the output of particles from particle scattering as we have seen.

The simple production rules that describe particle scattering in the Standard Model suggest a fundamental simplicity at the core of Reality that goes far beyond the speculations of philosophers and scientists of earlier ages.

# 5

# Quantum Turing Machines

## What are Turing Machines?
The linguistic view of the Standard Model leads to a number of questions. One important question is the nature of the Turing machine that accepts this language. A Turing machine is a generalized theoretical computer that is often used to analyze computational questions in computer science.

A personal computer can be viewed as a special purpose Turing machine. A personal computer has memory in the form of RAM and hard disks. A personal computer has built-in programs that tell it what to do when data is input into the computer. Similarly a Turing machine also has memory and instructions within it telling it how to handle input and how to produce output from a given input.

When we type input on a computer keyboard or have input come from another source such as the Internet or a data file, the input has to be in a form that the computer can handle. Similarly, the input for a Turing machine must have a specific form for the Turing machine to accept it, process it and then produce output. In the case of a Turing machine we say the input must be presented in a language that the Turing machine "accepts". In this context the word "accepts" means a format that the Turing machine can recognize and analyze so that it can process the data to produce output.

The language of the Standard Model is a type 0 computer language. A type 0 language requires a Turing machine to handle its productions. Because particle transitions are quantum and because the left side of a production rule can have several possible right sides (For example, a photon can transition to an electron-positron pair, a quark-antiquark pair and so on.) the Turing machine for the Standard Model language must be a non-deterministic Quantum Turing Machine.

## Features of Normal Turing Machines
Before examining a Quantum Turing Machine for the Standard Model we will look at the features of "normal" Turing machines. A normal Turing machine consists of a finitely describable black box (its features are describable in a finite number of statements) and an infinite tape. The tape plays the role of computer memory. The tape is divided into squares. Each square contains a symbol or character. The character can be the "blank" character or a symbol. A tape contains blank characters followed by a finite string of input symbols followed by blank characters.

The black box consists of a control part and a tape head. The control part has a finite set of rules built into it (the "program") and a finite memory that it uses as a scratch pad normally. The tape head can read symbols from the tape one at a time and can move the tape to the left, right, or not move it, based on instructions from the control part.

The tape head tells the control the symbols it is scanning from the tape and the control decides what action to have the tape perform based on the scanned symbols and information (the program and data) stored in the control's memory.

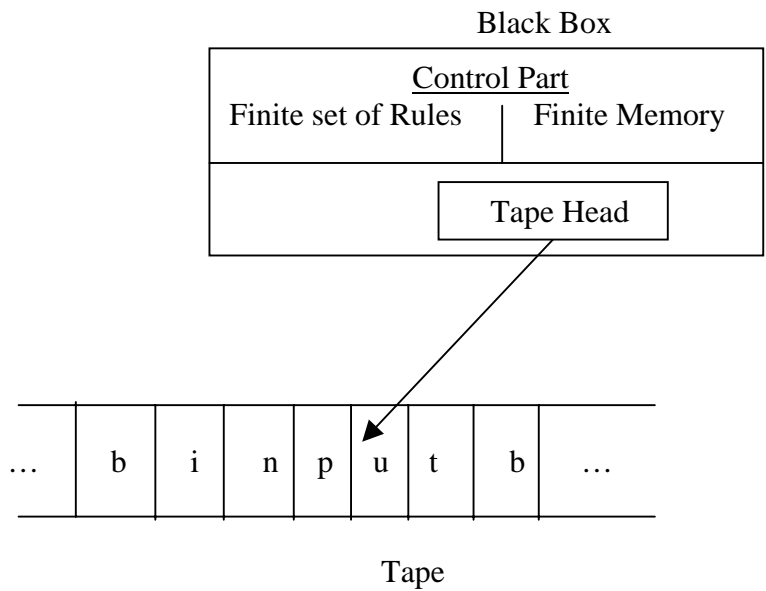

Figure. Schematic diagram of a Turing machine.

A set of input symbols is placed on the tape and the rules (program) in the control part are applied to produce an output set of symbols.

This process is analogous to elementary particle processes: an input set of particles interacts through various forces of nature and produces an output set of particles. The difference is that elementary particle processes are quantum probabilistic in nature. The laws of Physics (which appear to be finitely describable since they can be specified by the Standard Model lagrangian except for gravity) play the role of the finite set of rules.

**Quantum Probabilistic Grammars**
The major difference between Turing machine outputs and the outputs in particle physics are that output states in particle physics are quantum probabilistic. A given set of input symbols (particles) can produce a variety of output states with different probabilities calculable in the Standard Model. We need a Quantum Turing Machine to handle this more complex situation.

Quantum (and non-quantum) Turing Machines can be pictured in a convenient way by viewing the control part as containing a tape on which the rules are inscribed, the current state of the Turing machine is specified, and the current symbol being scanned by the tape head is stored.

The grammar rules of a Quantum Turing Machine are quantum probabilistic. In the simplest case each grammar rule has an associated number that we will call its relative probability amplitude. We will call this type of Quantum Probabilistic Grammar™ a *factorable Quantum Probabilistic Grammar*. The calculation of the probability for a transition from a specified input string to a specified output string is based on the following rules:

1. The relative probability amplitude for an input string to be transformed to a specified output string is the sum of relative probability amplitudes for each possible sequence of transitions that leads from the input string to the output string.

2. The relative probability amplitude for a sequence of grammar rule transitions is the product of the relative probability amplitudes of each transition.

3. The relative probability for an input string to be transformed to an output string is the absolute value squared of the relative probability amplitude for the input string to be transformed to the output string.

4. The absolute probability for an input string to be transformed to a specified output string is the relative probability for the input string to be transformed to the specified output string divided by the sum of the relative probabilities for the input string to be transformed to all possible output strings using the grammar rules. This rule guarantees the sum of the probabilities sums to one.

The $\phi^3$ theory example considered previously is an example of a factorable Quantum Probabilistic Grammar™. The grammar rules for the $\phi^3$ theory example were:

| Transition | Relative Probability Amplitude |
|---|---|
| $\phi \rightarrow \phi\phi$ | g |
| $\phi\phi \rightarrow \phi$ | g |

The relative probability amplitude for

input state: $\phi\phi$ → output state: $\phi\phi$

was shown to be

$$A_2(g) = \sum_{n=1}^{\infty} a_n\, g^{2n}$$

in the preceding chapter. The relative probability for this process is:

$$P_2 = |A_2(g)|^2$$

and the absolute probability for this process was shown in the preceding chapter to be

$$Q_2 = P_2/P$$

where

$$P = \sum_{n=1}^{\infty} P_n$$

with $P_n$ being the relative probability for a two $\phi$ particle input state to become an n $\phi$ particle output state.

A general Quantum Probabilistic Grammar is an *entangled Quantum Probabilistic Grammar* in which we cannot assign a probability amplitude to individual grammar rule transitions. The relative probability amplitude for a sequence of grammar rule transitions is not the product of the relative probability amplitudes of each transition.

Instead in an entangled Quantum Probabilistic Grammar the relative probability amplitude for a sequence of grammar rule transitions is a function of the combined set of grammar rule transitions.

The Standard Model grammar is in fact an entangled Quantum Probabilistic Grammar if we take account of the momenta and spins of the input and output particles. An illustration of the entanglement in the Standard Model is electron scattering through the exchange of two photons. The Feynman diagram for this process with particle momenta displayed is:

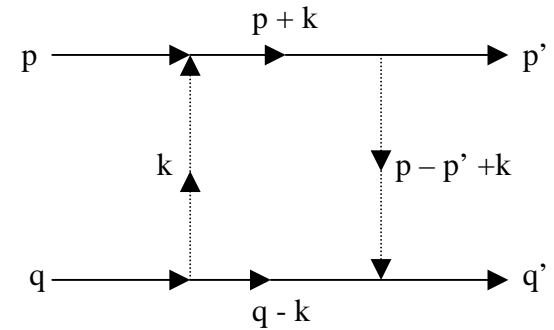

An electron with momentum p collides with an electron with momentum q and exchanges two photons (dotted lines). The momentum of each electron and photon is displayed. The outgoing electrons have the momenta p' and q'. The probability amplitude for this diagram has a factor of $e^4$ (a power of e for each of the four transitions or vertices that generate or absorb a photon). This factor is similar to the probability associated with a transition in factorable Quantum Grammars™. In addition there is a factor associated with the overall pattern of transitions. The resulting probability amplitude is proportional to

$$e^4 \int \frac{d^4k \, N(k)}{k^2[(p+k)^2 - m^2][(q-k)^2 - m^2][(p-p'+k)^2 - m^2]}$$

where N(k) is a numerator that depends on the electron momenta, spins, and the "loop" momentum k that is being integrated over. The integral depends on the details of the arrangement of the transitions and thus provides the entanglement. The calculation of probability amplitudes in Quantum Field theories is described in many books on the subject.

We conclude that the Quantum Probabilistic Grammar™ of the Standard Model is an entangled Quantum Probabilistic Grammar.

# 6

# The Standard Model Quantum Computer

**The Standard Model Quantum Turing Machine**
The Quantum Turing Machine that corresponds to the Standard Model has a number of exciting features that distinguish it from conventional Turing Machines.

First it accepts a language that has a finitely describable entangled Quantum Grammar™. Although the Standard Model has an entangled Quantum Grammar the grammar rules are finitely describable. Finitely describable means that the rules can be specified by a finite set of symbols. The rules generated from the interactions of the Standard Model are finite in number and each rule consists of a finite number of symbols. Thus the rules generated from the Standard Model are finitely describable.

A Quantum Turing Machine can be visualized as consisting of a control element and two tapes that play the role of computer memory. The control has a tape head that reads and writes symbols to the two tapes.

Tape I contains the specification of the grammar rules expressed as a finite string of symbols, the current state of the Quantum Turing Machine, and other data. Tape II contains the input string. After applying the grammar rules in a quantum probabilistic way an output state is generated. The output state is placed on Tape II in the simplest Quantum Turing Machine implementation.

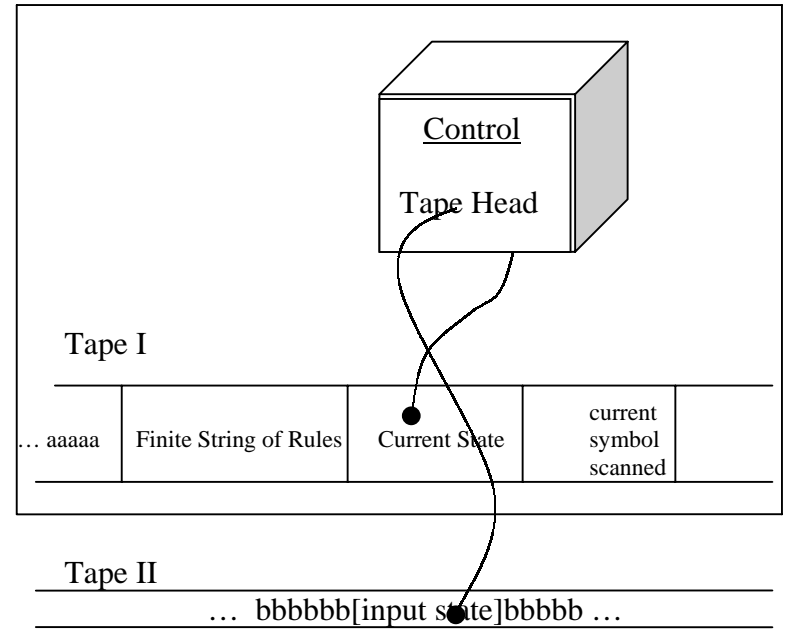

Figure. Quantum Turing Machine. Tape I plays the role of computer memory. Tape II is memory for input and output.

The behavior of a Quantum Turing Machine can be viewed as:

1. The Turing machine begins in the input state specified on tape II. An input string is placed on tape II. The other memory locations on tape II are filled with blank characters. In our case this string is a list of symbols for an input set of elementary particles that are about to interact. The Turing machine we are considering accepts any state consisting of a finite number of elementary particles. The connection between Turing machines and computer languages is brought out at this stage. A machine "accepts" a language if it can take any sentence (set of particles in our case) of the language, and perform a computation producing output (a set of output particles in our case). A Turing machine that accepts a language is an embodiment of the grammar of the language.

2. The Quantum Turing Machine applies the grammar rules to the input set of states in all possible ways to produce an output state that is a quantum superposition of states. Each possible output state has a certain probability of being produced.

3. The probability for producing a specified output state from a specified input state can be calculated as we illustrated in a simple example earlier using the relative probabilities associated with the Quantum Grammar™ rules of the modified $\phi^3$ theory.

4. The set of possible states of a Quantum Turing Machine[4] is infinite unlike non-Quantum Turing Machines that only have a finite number of possible states.

The Standard Model Quantum Turing Machine has some distinctive features:

1. Since the order of the particles in the input state string is not physically important we will consider the input string to be actually all permutations of the order of the particles in the input.

2. Since the Turing machine is quantum the rules are probabilistic in nature: a given set of input particles will in general produce many possible output particle states. Each output state will have a certain probability of being produced that can be calculated using the Standard Model.

3. The Quantum Grammar™ rules of the Standard Model Quantum Turing Machine have internal symmetries that result in symmetries in the input and output states.

4. Since the momenta and spins of the input and output particles are physically very important the Standard Model Quantum Turing Machine must take account of these properties in the input and output states as well as internally when calculating transition probabilities.

So we must picture the input particle state on tape II as containing not only the particle symbol but also momenta and spin data.

To get an idea of how a Quantum Turing Machine would take an input set of particles and produce a set of output particles we will consider the case of two electrons colliding with such energy that an electron-positron pair is created:

$$ee \rightarrow eepe$$

where e represents an electron and p represents a positron (the electron's antiparticle). One of the corresponding Feynman-like diagrams is:

---

[4] D. Deutsch, Proceedings of the Royal Society of London, A **400** 97 (1985) describes (universal) Quantum Computers and points out they can simulate continuous physical systems because they have a continuum (infinite number) of possible states. As page 107 points out "a quantum computer has an infinite-dimensional state space". Quantum Computers are equivalent to Quantum Turing Machines as we will see.

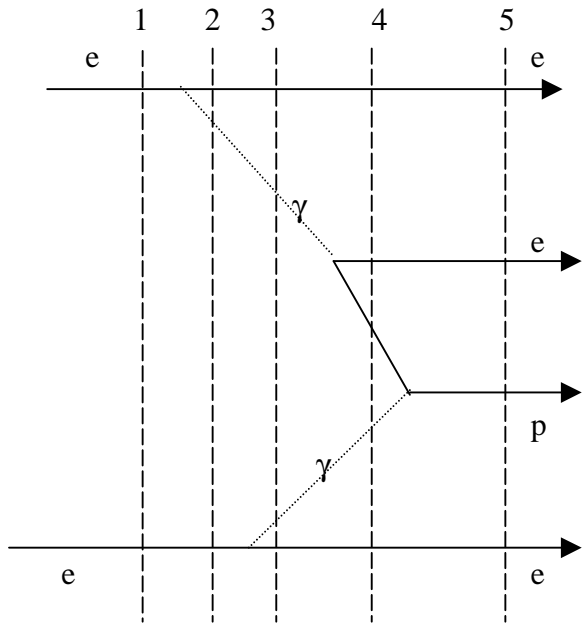

where γ represents a photon. The input string can change according to the grammar rules in the following way:

$$ee \rightarrow e\gamma e \rightarrow e\gamma\gamma e \rightarrow eep\gamma e \rightarrow eepe$$

Since the Quantum Turing Machine is probabilistic there are many – in fact an infinite number – of ways in which the transition

$$ee \rightarrow eepe$$

can take place – each with its own probability of happening. The sample sequence of transitions shown above is only one of these possible ways. The total probability of this transition is the square of the sum of the probability amplitudes for all possible ways according to quantum mechanics. Nature requires us to take account of all possible ways of transitioning from the input state of particles to the output state of particles. The total probability of the output state being produced is a sum of the contributions of all the possible alternate ways of reaching that output state.

In addition, electron-electron scattering can produce many other output states depending on the initial energy of the electrons. Each output state has its own probability of occurring. Some examples are:

$$ee \rightarrow eq\underline{q}e$$

$$ee \rightarrow eepepe$$

$$ee \rightarrow e\mu\underline{\mu}e$$

$$ee \rightarrow e\mu\underline{\mu}epe$$

where q represents a quark, $\underline{\mu}$ represents the muon antiparticle and $\underline{q}$ represents an antiquark.

The Quantum Turing Machine representation does raise several interesting prospects for the theory of elementary particles embodied in the Standard Model. First, the Quantum Turing Machine representation raises the possibility that some of the powerful techniques and general results of the theory of computation can be brought over to physics and perhaps provide guidance on the next stage after the Standard Model.

Secondly, and perhaps more importantly, the separation of the input and output states (they are on tape II) from the intermediate calculational states of the Turing machine (that are on tape I) is suggestive of a somewhat different approach to the fundamentals of particle interactions: The space-time of the incoming and outgoing particles may be different from the "space-time" describing the interactions and internal structure of the interacting particles. This view is based on thinking of tapes I and II as representing separate space-times.

A precursor of this point of view appears in Quantum Field Theory. In Quantum Field Theory the interaction of particles is viewed as consisting of three phases: an initial state where the particles are widely separated and distinct, an interaction region where the particles "collide" and interact perhaps creating new particles, and a final state where the outgoing particles are widely separated.

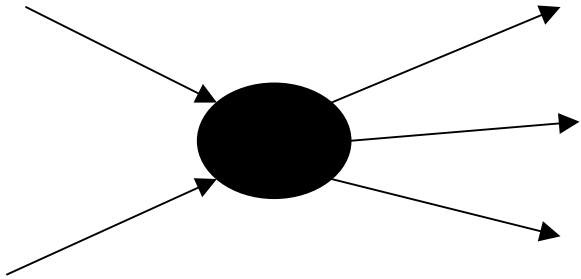

Figure. Two particles collide and generate a three particle outgoing state in Quantum Field Theory.

In conventional Quantum Field Theory the space-time in the interacting region is conventionally assumed to be the same as the space-time of the incoming and outgoing states. Nevertheless Quantum Field Theory distinguishes the interaction region from the region of the incoming and outgoing particles.

The SuperString approach to the theory of elementary particles introduces a separate space-time to describe the elementary particles. Elementary particles are viewed as strings vibrating in this space-time. Can one view the SuperString space-time as tape I and the external behavior of the elementary particles taking place in normal space-time as tape II? Perhaps the Quantum Turing Machine representation of the Standard Model is the key to the next level of our understanding of elementary particles and Nature.

# 7

# Quantum Computers and Fock Space

Up to this point we have been looking at Quantum Turing Machines. Recently much excitement has been generated by Quantum Computers. A Quantum Computer is an alternate and earlier formulation of the Quantum Turing Machine concept. In 1982 Richard Feynman[5] described a new theoretical concept called a Quantum Computer. Quantum Computer concepts had been developed by Benioff[6], Deutsch and others in preceding years.

A Quantum Computer can be viewed as similar to a "normal" computer in many ways. However it is not deterministic. It transitions from one state to another in a quantum probabilistic way.

---

[5] R. P. Feynman, International Journal of Theoretical Physics, **21**, 467 (1982).
[6] Paul Benioff, Jour. Stat. Phys. Vol 22, p.563 (1980); Jour Stat. Phys. Vol 29, p 515 (1982); Phys. Rev. Letters Vol 48, p 1581 (1982); Int. Jour Theoret. Phys. Vol 21, 177 (1982).

Feynman's Quantum Computer was particularly adapted to simulating physical quantum systems. He gave a concrete picture of it as a space-time lattice that has two possible states at each point on the space-time lattice.

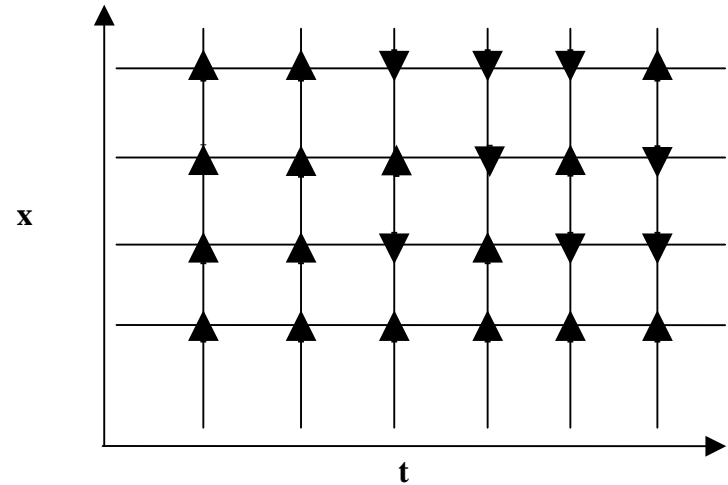

Figure. A Quantum Computer viewed as a space-time lattice of points. At each intersection point there are two possible states: occupied or unoccupied. Alternately we can use a spin ½ description and say that we have spin "up" or spin "down" at each space-time point (indicated with up arrows or down arrows). Or we can say that a computer bit exists at each space-time point that can have the value 1 or 0.

Feynman then hypothesized that every finite quantum mechanical system could be exactly described (exactly simulated) with such a system if the system were treated as a finite lattice of interacting spins and a suitable interaction was "chosen" between the spins at the adjacent lattice points. (We are interested in infinite lattices.)

The Quantum Computer would start with the spins on the lattice set up to correspond to an initial quantum state of the quantum system being simulated. Then the interaction between the spins would cause transitions in the spins mirroring the evolution of the quantum system being modeled. The final state of the Quantum Computer would correspond to the final state of the quantum system being simulated. In particular the probability of a transition to a specified output state of the Quantum Computer would be identical to the probability of a transition of the quantum system being modeled to its corresponding output state.

Feynman's concept of a Quantum Computer was concrete and physical – you could view it as a lattice of spins – an idea that physicists were comfortable with. Since then, numerous individuals have explored theoretical aspects of quantum computers including D. Deutsch.[7] Deutsch provided a more general formulation of Quantum Computers based on quantum bits which he called a universal quantum computer. His universal quantum computer is capable of "perfectly simulating every finite, realizable physical system." Recently, quantum computers have been shown to offer significant advantages over normal computers in some types of calculations[8].

The concept of the Quantum Computer as developed by Feynman and others is an alternate representation of the Quantum Turing Machine concept. The Quantum Turing Machine concept is based on a linguistic approach with a Quantum Grammar, and a vocabulary of symbols. We chose to develop this representation of Quantum Computers because it was the natural approach for the Standard Model. The Standard Model interaction lagrangian specified the Quantum Grammar with a vocabulary consisting of the fundamental particles of the Standard Model.

Can we map our Quantum Turing Machines to Quantum Computers and show they are equivalent concepts? Actually this mapping is quite easy to do.

The mapping or correspondence between the Quantum Turing Machine and a Quantum Computer of the type of Feynman and Deutsch is based on the following correspondences:

$$\text{Symbol} \rightarrow \text{number} \rightarrow \text{bit pattern} \rightarrow \text{lattice spins}$$

Each symbol in the vocabulary of a Quantum Turing Machine can be given a numerical value just as alphabetic characters in a real computer are given numerical values internally. The numeric value of each symbol can be written in base 2 as a binary string of zeros and ones just as modern computers do. The string of zeros and ones specifies bits in a Quantum Computer. The initial quantum state of a Quantum Computer is the combined set of bits corresponding to the symbols in the input string of symbols of the corresponding Quantum Turing Machine.

The Quantum Grammar of the Quantum Turing Machine maps to an interaction between the spins on the lattice of the Quantum Computer. The interaction between the spins must correspond to the Quantum Grammar™ in the sense that, for any given input state, the output state of the Quantum Computer must correspond to the equivalent output state of the Quantum Turing Machine.

Another way of stating this correspondence between the Quantum Grammar™ and the lattice spin interaction is that the probability of producing a specified output state from a specified input state must be the same for a Quantum Turing Machine and its equivalent Quantum Computer. The Quantum Grammar must correspond to an equivalent interaction between the spins on the Quantum Computer lattice.

$$\text{Quantum Grammar} \leftrightarrow \text{Interaction between spins of lattice}$$

Thus we have established a mapping between Quantum Turing Machines and Quantum Computers.

---

[7] D. Deutsch, Proceedings of the Royal Society of London, A **400** 97 (1985) and A **425** 73 (1989). P. Benioff explored state descriptions even earlier. See footnote 6.
[8] P. W. Shor, Algorithms for Quantum Computation: Discrete log and Factoring, Bell Laboratories paper, 1998. Many other excellent papers on the advantages of Quantum Computers have appeared since Shor's seminal paper.

## The Continuum Limit of a Quantum Computer – A Superspace

A Quantum Computer has a lattice of points. At each point there is a bit that can be in an "up" or a "down" state. Let us imagine "shrinking" the lattice so the space between lattice points becomes smaller and smaller. Eventually the lattice approaches what could be called the continuum limit where the points form a continuous space. At each point of this space there is a bit that can be either "up" or "down" and, as a result, we can view this space as a *superspace* similar to that encountered in Supersymmetry theories. We will call this limiting case of the Quantum Computer a *Continuum Quantum Computer*™.

At each lattice point the spin is either up or down. As the lattice spacing shrinks we encounter discontinuities in the spin. The discontinuities form a countable set of measure zero, i.e. they can be counted using the set of integers. We smooth over these discontinuities to have smoothly varying spins in the continuum limit of the lattice. The information contained in the smoothly varying spins is the evolving data in the Quantum Computer.

In Supersymmetry theories[9] there is a parameter space consisting of space-time coordinates (parameters) which we will denote as $x^\mu$, and Grassmann variable parameters denoted $\theta$ and $\overline{\theta}$. Grassmann parameters are not normal variables because they satisfy anti-commutation rules.

The Continuum Quantum Computer also has an extended space-time. It has a space-time parameter $x^\mu$ together with a spin at every space-time point. The index $\mu$ is a number specifying coordinate. For ordinary space-time $\mu$ ranges from 0 to 3. The coordinate $x^0$ is the time coordinate. The coordinates $x^1$, $x^2$ and $x^3$ are the space coordinates.

---

[9] We will follow the approach of D. Bailin and A. Love, *Supersymmetric Gauge Field Theory and String Theory* (Institute of Physics Publishing, Philadelphia, PA, 1994) pages 13, 23, and 34.

The spin "dimension" parameter of the Continuum Quantum Computer can be viewed as a Grassmann variable parameter. To illustrate this correspondence let us imagine that we construct a quantum wave function $\psi(x)$ that specifies the probability amplitude to find "spin up" vs. "spin down" at each space-time point. Let

$$\psi(x) = \begin{pmatrix} \psi_1(x) \\ \psi_2(x) \end{pmatrix}$$

be the (Weyl) spinor wave function for the quantum computer in the continuum limit. The component $\psi_1(x)$ is the probability amplitude that the spin is "up" at the point x, and the component $\psi_2(x)$ is the probability amplitude that the spin is "down" at the point x. We assume $\psi(x)$ is continuous except possibly for a set of measure 0.

$\psi(x)$ can be rewritten in terms of spinors as:

$$\psi(x) = \psi_1(x)s_1 + \psi_2(x)s_2$$

where

$$s_1 = \begin{pmatrix} 1 \\ 0 \end{pmatrix}$$

and

$$s_2 = \begin{pmatrix} 0 \\ 1 \end{pmatrix}$$

are spinors. Now the Grassmann variable $\theta$ has a Weyl spinor index $\theta_a$ as well. So we can represent our wave function as

$$\psi(x) = \psi_1(x)\theta_1 + \psi_2(x)\theta_2$$

Thus our Continuum Quantum Computer maps directly to a space of the type required for Supersymmetry.

Now we take the Supersymmetric transformations and reinterpret them in terms of the physics of the Continuum Quantum Computer. In a Supersymmetric space the form of a Supersymmetric "rotation" is:

$$x^\mu \rightarrow x^\mu + a^\mu - i\xi\sigma^\mu\overline{\theta} + i\theta\sigma^\mu\overline{\xi}$$

$$\theta \rightarrow \theta + \zeta$$

$$\overline{\theta} \rightarrow \overline{\theta} + \overline{\zeta}$$

where $a^\mu$ specifies a shift (translation) in space-time coordinates, and $\zeta$ and $\overline{\zeta}$ are constant anti-commuting Grassmann parameters. A Supersymmetric "rotation" can shift the space-time coordinates and inter-mix them with the Grassmann parameters. The $-i\xi\sigma^\mu\overline{\theta} + i\theta\sigma^\mu\overline{\xi}$ expression is an ordinary numeric expression – not an anti-commuting Grassmann value. The Supersymmetric "rotation" is reminiscent of the rotation between space and time that we see in the Theory of Special Relativity. Only here, in Supersymmetric space, it rotates between ordinary space-time coordinates and Grassmann coordinates.

The interpretation of the Supersymmetric "rotation" from the viewpoint of the Continuum Quantum Computer™ is quite interesting. To understand it we must first realize that the information in the memory of a Quantum Computer is not only the bits in the memory lattice. The distribution of the bits – how they change as one goes through memory is also part of the information contained in the Quantum Computer's memory. With that in mind we can view the Supersymmetric transformation that intermixes space-time and the Grassmann spinor parameters as a rotation between the data bits in the Continuum Quantum Computer and the space-time of the Continuum Quantum Computer.

If the spin interactions of the Continuum Quantum Computer™ are invariant (unchanged) under a Supersymmetric transformation then the transformed Continuum Quantum Computer is equivalent to the original Continuum Quantum Computer. (We assume that edge effects – effects of the edge of the Continuum Quantum Computer's space are negligible.)

## SuperSymmetric Continuum Quantum Computer

If we now imagine creating a Continuum Quantum Computer that is equivalent to a Quantum Turing Machine for an enhanced theory of elementary particles that includes SuperSymmetry, then its space will be of infinite extent. We will call this computer the SuperSymmetric Continuum Quantum Computer. (We can also create a Continuum Quantum Computer for SuperString theory as we show in a later chapter.)

The entire universe can be viewed as contained in the SuperSymmetric Continuum Quantum Computer™. The lagrangian for that theory can be added to the "edge" of the universe as the rules (or program) part of the quantum computer forming an augmented universe.

The lagrangian for the universe that we add to the "real" universe can be thought of as occupying a universe of its own. Spaces or universes of lagrangians have been discussed by Kenneth Wilson and others. An interesting question raised by the existence of a space of lagrangians is whether a principle exists in the space of lagrangians that can be used to deduce the lagrangian that Nature seems to implement just like there is a way to deduce laws of Physics from the Standard Model lagrangian. One possibility suggested by the computer framework of the discussion is a lagrangian that is generated through a self-organizing process of the sort envisioned in Artificial Intelligence studies. This possibility remains to be explored.

This representation of the enhanced SuperSymmetric Continuum Quantum Computer effectively reduces Nature to Language. Everything in the universe including the laws of Physics becomes part of a quantum computer. The space of the quantum computer is filled with bit patterns describing the data and "program" or rules of the computer.

## Group Symmetries of Quantum Computers

It is interesting to note that the type of symmetries of Quantum Computers that we have been discussing, and their implications for computation, has not been addressed in the literature on Quantum Computers.

## Particles and Physics Laws Become Bit Patterns

The SuperSymmetric Continuum Quantum Computer contains a space of bits that can describe particles. It also contains bit patterns representing the interactions of particles. Both the particles in the space and the lagrangian possess a common symmetry. The lagrangian is the program and the space is the memory of the computer. Together they form one integral whole.

The SuperSymmetric Continuum Quantum Computer adds bits representing the internal symmetries to each space-time point just as the Standard Model implements internal symmetries directly without "deriving" them from a more fundamental concept such as a space-time with many extra dimensions beyond the 4-dimensional space-time of our experience that collapses to 4-dimensional space-time generating the internal symmetries through a Kaluza-Klein mechanism. So we can view the SuperSymmetric Continuum Quantum Computer as the limit of a space-time lattice with a spin at each intersection point and a dangling thread of bits:

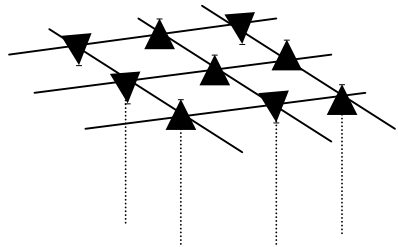

Figure. Lattice representation of SuperSymmetric Continuum Quantum Computer with bits for the internal symmetries. The Continuum Quantum Computer emerges as the limit when the lattice spacing decreases to zero.

Only 6 bits are actually required for the tightest, most minimal representation of approximately 36+ particles. The actual number of bits and the role of each bit could be chosen to provide the most elegant representation of a particle's internal quantum numbers based on the SuperSymmetric symmetries.

The space of the lattice has been tacitly discussed as if it was coordinate space. Actually we shall see in a few sections that the continuum limit of a Quantum Computer (where the lattice spacing shrinks to zero) can be treated as a second-quantized Fermi field. The natural interpretation of the lattice space of a Quantum Computer is a "momentum" space.

So the particles' momentum values and internal quantum numbers are the data resident in the Quantum Computer. The momentum distribution of a particle can be Fourier-transformed to a coordinate space representation.

## Reality is Reduced to Language

Having seen particles first become symbols and then transform to bit patterns in a Continuum Quantum Computer™, having seen the interactions between bit patterns similarly reduced to bit patterns we now come to the conclusion that the universe and its physical laws – Reality – is in essence is linguistic – Language.

## The Meaning of the Standard Model Language

The purposes of language include understanding and communication. A language also reveals something of the thinking patterns of the speakers of the language. Language not only expresses our thoughts but it also helps shape them. What can one make of the linguistic approach to the Standard Model? What does the language of the Standard Model tell us about Nature? What are its capabilities for expressing ideas? Where is the Rosetta stone that could map the language of the Standard Model to a form that we can understand in terms of its purpose and its goals? What are the reasons for its form?

The elementary particles that appear in the Standard Model and its relatively simple form is amazing considering the possibility of a universe with untold numbers of unrelated particles each described by a different theory. Many physicists complain the Standard Model has too many particles. But actually it is a small miracle that a theory containing a few particles (36+), and describable by a lagrangian with a few parts, accounts for all known phenomena involving matter and radiation in the universe (except gravity).

This being so, one has to ask if there is a human-understandable interpretation of the theory at a level that can be explained in a few words understandable to an informed layman. (Feynman once said any good theory should have a simple idea behind it.)

John A. Wheeler developed a similar question in one of his papers[10]: "An unfamiliar computer from far away stands at the center of the exhibition hall. Some of the onlookers marvel at its unprecedented power; others gather in animated knots trying, but so far in vain, to make out its philosophy, its logic, and its architecture. *The central idea of the new device escapes them. The central idea of the universe escapes us.* [italics added]" What is the central idea behind the language of the Standard Model Quantum Computer and thus the material universe? Have we not found Wheeler's "unfamiliar computer from far away" to be our universe as described by the Standard Model?

**Computer Languages and the Standard Model Language**
One way to begin to develop an understanding of the central idea of the universe is to first explore the more limited possibility of deducing the purpose of a computer language from its features. We wish to take a language and deduce its purpose and goals from the structure and capabilities of the language. Then we would like to apply this approach to the Standard Model computer language to develop a deeper understanding of the universe.

---
[10] J. A. Wheeler, The Computer and the Universe, International Journal of Theoretical Physics, **21,** 557 (1982).

A reasonable starting point is to consider the set of computer languages with which we are familiar: COBOL, FORTRAN, C, C++, Java, and so on. Given one of these languages could we deduce its purpose, objectives and "philosophy"? Imagine yourself an alien and someone hands you a description of, say, the C++ language describing the features of C++ programming (written in your alien language). Could you deduce the purpose and "philosophy" of this language? An interesting question.

The C++ language is used for mathematical calculations in science, engineering and other practical applications. It is also used to manage devices such as printers, other computer devices and industrial machinery. C++ implements a "philosophy" called Object Oriented Programming. Could we have deduced this "philosophy" from the structure of the C++ language? Personally, I think not – not without serious effort – and not without an understanding of the drawbacks of the language that preceded it – the C programming language.

The C language was excellent for small scale programming projects. But because it required the programmer to track every variable and all the details of a program the C language was not appropriate for large programming projects. The C++ language was developed to handle large-scale programming. It implemented a design philosophy called Object Oriented Programming. This approach was based on grouping variables into objects. Objects were used as the central entities in programming and they were manipulated in C++ programs. By grouping many variables within an object and then manipulating the objects C++ programs are easier to create, and easier to maintain and modify.

A casual examination of C++ language features would not reveal these concepts and purposes. A detailed study of the C++ language might lead an imaginative alien to the rationale for the object-oriented features of C++.

Returning to physics, the example of the C++ language suggests that we might understand the language of the Standard Model and its "philosophy" by understanding it within the context of other languages, and the difficulties with other languages that the Standard Model language resolves. *This suggestion argues for a meta-theory of physics theories*: a theory of the nature of physics theories and of their comparative features – a Theory of Theories as we discussed earlier.

## Equivalence of Continuum Quantum Computers and Second-Quantized Fermi Fields – Fock Space

The Continuum Quantum Computer™ can also be viewed as defining a Fock space that leads to a second-quantized Fermi field representation. The formalism of second-quantized Fermi fields can be used to describe Quantum Computers in the continuum limit. They can also approximate Quantum Computers defined by a discrete lattice.

A Quantum Computer can be described[11] as a processor consisting of a set of M 2-state (up or down) observables denoted $\{n_i\}$ where i ranges from 0 through M – 1, and an infinite memory consisting of a set of 2-state observables denoted $\{m_i\}$ where i ranges from -∞ to ∞. (∞ is the symbol for infinity.) The computer's memory can be viewed as either linear like a tape or arranged in a lattice that is numbered linearly. The tape head is assumed to be "reading" at a memory location k on the tape.

The state of a Quantum Computer can be specified with a basis of state vectors of the form:

$$| n_0, n_1, n_2, \ldots, n_{M-1}; \ldots m_{-1}, m_0, m_1, \ldots > | k >$$

or, in shorthand notation,

$$| \mathbf{n}; \mathbf{m} > | k >$$

The vector $| k >$ specifies the state of the tape head. These vectors are eigenvectors of the observables and span the space of states of the Quantum Computer.

Each 2-state observable $n_i$ or $m_i$ can have the values 0 or 1 (bit "off" or bit "on"). This approach is equivalent to the spin ½ picture where the values are –½ or +½.

The basis states can be easily interpreted as a set of multi-particle states in which each 2-state observable corresponds to a quantum level that can contain zero particles or one particle.[12] Thus we can view the basis of Quantum Computer states as a number representation for fermions (combined with the tape head state).

In the discussion that follows we will begin by lumping the 2-state observables of the processor together with those of the computer tape for the sake of simplicity. (We will ignore the tape head for the moment.) The discussion will then treat these sets of observables separately.

The vacuum state for the number representation is the vector with all 2-state observable values 0:

$$\Phi_V = | 0, 0, 0, \ldots, 0; \ldots 0, 0, 0, \ldots >$$

or, in short,

$$\Phi_V = | \mathbf{0}; \mathbf{0} >$$

Creation and destruction operators can be defined that create and connect multi-particle states (states with one-bits). The operator $d_i^\dagger$ creates a particle in the i$^{th}$ quantum state or, in other words, changes the i$^{th}$ 2-state observable bit from zero to one.

$$\begin{array}{c} \text{i}^{th} \text{ quantum state} \\ \text{(2-state observable)} \\ \downarrow \\ d_i^\dagger \Phi_V = | 0, 0, \ldots, 1, \ldots 0, 0, 0, \ldots > \end{array}$$

---

The operator $d_i$ destroys a particle in the $i^{th}$ quantum state or, in other words, changes the $i^{th}$ 2-state observable bit to zero from one. Applying a destruction operator to the vacuum state annihilates it:

$$d_i \Phi_V = 0$$

In order to guarantee the 2-state nature of the quantum levels (observables) we impose anti-commutation relations on $d_i^\dagger$ and $d_i$.

$$\{ d_i, d_i^\dagger \} = d_i d_i^\dagger + d_i^\dagger d_i = 1$$

$$\{ d_i, d_i \} = 0$$

$$\{ d_i^\dagger, d_i^\dagger \} = 0$$

These anticommutation relations guarantee the spectrum of each observable will be one or zero (See Bjorken and Drell for the details). The relation of $d_i$ and $d_j$ where i and j are different is not as clear until we realize that the states

$$d_i^\dagger d_j^\dagger \Phi_V = c d_j^\dagger d_i^\dagger \Phi_V$$

for $i \neq j$ where c is a phase factor ($e^{i\phi}$). The order in which bits are set cannot influence the state that is created except for an irrelevant phase factor. (It is irrelevant because probabilities are calculated from the *absolute* value squared of amplitudes.) The phase factor can be chosen to be –1 to obtain consistency with the anticommutation relations above. Similarly

$$d_i d_j \Phi = c' d_j d_i \Phi$$

for $i \neq j$ where c' is a phase factor ($e^{i\phi}$). The phase factor can again be chosen to be –1 to obtain consistency with the anticommutation relations above.

Pulling these considerations together we obtain the anticommutation relations:

$$\{ d_i, d_j^\dagger \} = \delta_{ij}$$

$$\{ d_i, d_j \} = 0$$

$$\{ d_i^\dagger, d_j^\dagger \} = 0$$

where $\delta_{ij}$ equals 1 if i = j and $\delta_{ij}$ equals zero otherwise.

The general Quantum Computer state can now be represented by a multi-particle state:

$$| n_0, n_1, \ldots, n_{M-1}; \ldots m_{-1}, m_0, m_1, \ldots > | k > = N d_i^\dagger d_j^\dagger \ldots d_q^\dagger d_p^\dagger \Phi_V | k >$$

where N is a normalization constant. While it would be possible to go on and discuss normalizing states and other issues of the number representation we will refer the reader to Bjorken and Drell for further details.

The aspect that we would like to explore further is the introduction of a second-quantized Fermi field to characterize a Quantum Computer. For this purpose we define the field

$$\psi(z, t) = \sum_i u_i(z, t) d_i$$

where the sum is over the lattice sites labeled with index number i. The Hermitean conjugate field is

$$\psi^*(z, t) = \sum_i u_i^*(z, t) d_i^\dagger$$

The operators $\psi$ and $\psi^*$ satisfy the anti-commutation relations

$$\{\psi^*(z, t), \psi^*(z', t)\} = 0$$

$$\{\psi(z, t), \psi(z', t)\} = 0$$

$$\{\psi(z, t), \psi^*(z', t)\} = \sum_i u_i(z, t) u_i^*(z', t) = \delta(z - z')$$

where the set of functions $u_i(z, t)$ form a complete orthonormal set of functions. These functions should be solutions of an equation that describes the time evolution of the quantum computer (perhaps similar to the Schrodinger equation of Quantum Mechanics). They satisfy the orthonormality condition:

$$\int dz\, u_i(z, t) u_j^*(z, t) = \delta_{ij}$$

The parameter z can be treated above as a one-dimensional spatial parameter.

In the continuum limit the index i can be transformed into a continuous variable that can be viewed as the momentum conjugate to the z parameter.

Thus we have arrived at a second-quantized Fermi field $\psi(z, t)$ formulation that describes a Quantum Computer.

A one-particle state can be specified by:

$$\int dz\, u_i(z, t) \psi^*(z, t) \Phi_V = d_i^\dagger \Phi_V = |\, 0, 0, \ldots, 0, \overset{i^{th}}{1}, 0, \ldots >$$

Multi-particle states can be constructed through a straightforward generalization.

The non-zero anti-commutation relation between $\psi$ and $\psi^*$ reflects a measurability issue for Quantum Computers that is familiar from the analogous phenomena in the theory of Quantum Fields. If we attempt to measure a field at two points that can be connected by a light signal then an uncertainty principle exists that prevents the exact measurement of the field at each point.[13] In the case of a Quantum Computer the measurement of a quantum bit(s) at one point in a Quantum Computer and the measurement of another bit in another part of the computer that can be connected to it by a light signal are incompatible – they satisfy an uncertainty principle that prevents their simultaneous measurement.

Having arrived at a second-quantized Fermi field description of a Quantum Computer it is interesting to raise the question of how it can be programmed from the perspective of this representation. Programming a Quantum Computer requires the specification of a quantum dynamics that describes the evolution of the Quantum Computer from an initial state.

There are several approaches that appear reasonable at first glance:

1. Specify a multi-particle (multi-bit) Hamiltonian using a Schrodinger-like equation that describes the time evolution of the particles (bits) of the Quantum Computer. This approach is limited by the fact that the number of particles (bits) is fixed – unlike most computations where the number of particles (on bits) varies as the computation progresses.

2. Define a second-quantized field theory for $\psi$ in z space with interactions that specify a computation. The interactions can appear in a variety of forms. Renormalization questions (infinities that appear in computations) are not an issue since the discreteness of the Quantum computer (the lattice spacing) provides a natural cutoff eliminating problems with infinities. Interaction terms that could appear in a second-quantized field theory include:

$$a_1 \psi^* \psi \psi^* \psi + a_2 \psi^* \frac{\partial \psi}{\partial z} \psi^* \frac{\partial \psi}{\partial z} + \ldots$$

This approach has the difficulty that it is difficult to map the interaction terms into bit operations in a direct way making it difficult to specify a program for a Quantum Computer in this manner.

---

[13] N. Bohr and L. Rosenfeld, Kgl. Danske Videnskab. Selskab. Mat.-Fys. Medd., **12**, 8 (1933) and Phys. Rev. **78**, 794 (1959).

3. Directly use the creation and annihilation operators to define a program for a Quantum Computer. A simple example of this approach is to consider a Quantum Computer that embodies the dynamics of an "on" bit "moving" on a tape. We assume a set of simple time-independent Hamiltonian terms implement the dynamics:

$$H = \sum_i d_{i+1}^\dagger d_i$$

Let us assume the Quantum Computer starts in a state with only the $0^{th}$ bit on (or in other words a particle at the $0^{th}$ lattice or tape position). Let us also define the set of one particle states:

$$\Phi_i = d_i^\dagger \Phi_V$$

Then with the time evolution operator:

$$S = e^{iHt}$$

we obtain the state of the Quantum Computer (initially in state $\Phi_0$) at time t:

$$\Phi(t) = S\Phi_0$$

or in expanded form:

$$\Phi(t) = \sum_{n=0}^\infty \frac{(it)^n}{n!} \left(\sum_j d_{j+1}^\dagger d_j\right)^n \Phi_0$$

$$= \sum_{n=0}^\infty \frac{(it)^n}{n!} \Phi_n$$

The state of the Quantum Computer has evolved into a superposition of one-particle states residing at memory positions 0, 1, 2, … (or alternately a superposition of Quantum Computer states with one bit "on").

The preceding example illustrates a Quantum Computer programmed to evolve dynamically according to a specified Hamiltonian. More realistic, and therefore more complex, programming can be done with this approach using creation and annihilation operators.

The preceding example illustrates the case of a nearest neighbor interaction – an interaction between adjoining bits or tape locations. The Quantum Computer state evolves into a state composed of a superposition of one "on" bit (one-particle states) states.

We have just seen the mapping of a Quantum Computer to a many-particle formalism and to a second-quantized Fermi field without differentiating between the processor and the computer memory and neglecting the positioning of the tape head. These complications can easily be handled within the framework of the mapping.

As pointed out earlier, the general state of a Quantum Computer can be symbolized by

$$| n_0, n_1, n_2, \ldots, n_{M-1}; \ldots m_{-1}, m_0, m_1, \ldots \rangle | k \rangle$$

We can define creation and annihilation operators for the memory bits $m_i$ as before:

$$\{d_i, d_j^\dagger\} = \delta_{ij}$$

$$\{d_i, d_j\} = 0$$

$$\{d_i^\dagger, d_j^\dagger\} = 0$$

where $\delta_{ij}$ equals 1 if i = j and $\delta_{ij}$ equals zero otherwise. In addition we now define creation and annihilation operators for the processor bits which we will denote $a_i$ having similar anticommutation relations:

$$\{ a_i, a_j^\dagger \} = \delta_{ij}$$

$$\{ a_i, a_j \} = 0$$

$$\{ a_i^\dagger, a_j^\dagger \} = 0$$

Lastly we define a creation operator $c^\dagger$ and an annihilation operator $c$ corresponding to the tape position observable. Assuming the tape is semi-infinite – the tape locations range from 0 to $+\infty$. Therefore the spectrum of this observable – the tape read position – ranges from 0 to $+\infty$, and the creation and annihilation operators should be defined to satisfy Bose commutation relations:

$$[ c, c^\dagger ] = 1$$

$$[ c, c ] = 0$$

$$[ c^\dagger, c^\dagger ] = 0$$

With these additional operators we can now define a Quantum Computer state as:

$$| n_0, n_1, \ldots, n_{M-1}; \ldots m_{-1}, m_0, m_1, \ldots > | k > =$$

$$N (c^\dagger)^k a_r^\dagger a_s^\dagger \ldots a_t^\dagger a_u^\dagger d_i^\dagger d_j^\dagger \ldots d_q^\dagger d_p^\dagger \Phi_V$$

where $\Phi_V$ is the vacuum state of the Quantum Computer and N is a normalization constant.

Following the same strategy as earlier it is possible to define fields

$$\psi(z, t) = \sum_i u_i(z, t) d_i$$

and

$$\phi(z, t) = \sum_j w_j(z, t) a_j$$

where the sum is over the $\psi$ lattice sites labeled with index number i and for $\phi$ over the $\phi$ lattice sites labeled with index number j. The $\psi$ and $\phi$ lattices are of course different. The Hermitean conjugate fields are

$$\psi^*(z, t) = \sum_i u_i^*(z, t) d_i^\dagger$$

and

$$\phi^*(z, t) = \sum_j w_j^*(z, t) a_j^\dagger$$

The operators $\psi$ and $\psi^*$, and $\phi$ and $\phi^*$, satisfy the anti-commutation relations

$$\{ \psi^*(z, t), \psi^*(z', t) \} = 0$$

$$\{ \psi(z, t), \psi(z', t) \} = 0$$

$$\{ \psi(z, t), \psi^*(z', t) \} = \sum_i u_i(z, t) u_i^*(z', t) = \delta(z - z')$$

$$\{ \phi^*(z, t), \phi^*(z', t) \} = 0$$

$$\{ \phi(z, t), \phi(z', t) \} = 0$$

$$\{ \phi(z, t), \phi^*(z', t) \} = \sum_j w_j(z, t) w_j(z', t) = \delta(z - z')$$

where the set of functions $u_i(z, t)$ and $w_j(z, t)$ form complete orthonormal sets of functions. In the case of the processor bits (2-state observables or particles) the number of observables is assumed to be infinite (although the number of occupied states is finite.) These functions $\psi$ and $\phi$ should be solutions of equations that describe the time evolution of the quantum computer processor and tape. Normally one would expect the processor and tape states to be interconnected as they evolve with time.

The parameter z has been treated above as a one-dimensional spatial parameter. It could have been a two-dimensional, three-dimensional, etc. parameter. We could have started with a multi-dimensional tape or a set of multidimensional tapes – a *Polycephalic Quantum Computer*™.

The tape head observable with creation and annihilation operators c and $c^\dagger$ should not be mapped to a corresponding field since there is only one observable in the case presented. This situation can however be changed by introducing an array (lattice) of tape heads. Classical polycephalic Turing machines with multiple tape heads have been discussed in the literature. These Turing machines had a finite number of tape heads. We can extend the number of tape heads to be infinite in number (and perhaps form a multidimensional lattice) in the case of Quantum Computers. This extension enables us to introduce a second-quantized Bose field corresponding to the lattice of tape heads:

$$\sigma(z, t) = \sum_j s_j(z, t) c_j$$

$$\sigma^*(z, t) = \sum_j s_j^*(z, t) c_j^\dagger$$

The Bose operators $\sigma$ and $\sigma^*$ satisfy the commutation relations

$$[\sigma^*(z, t), \sigma^*(z', t)] = 0$$

$$[\sigma(z, t), \sigma(z', t)] = 0$$

$$[\sigma(z, t), \sigma^*(z', t)] = \sum_j s_j(z, t) s_j^*(z', t) = \delta(z - z')$$

The appearance of both Fermi and Bose fields in Quantum Computers, and the apparent similarity of the lattice space of the Quantum Computer memory to a Supersymmetric space that could support Supersymmetry, leads us to consider the possibility that a Quantum Computer formalism may provide a foundation for an ad hoc development of SuperStrings. This possibility will be considered in a few chapters.

**Purpose of the Standard Model Language**

The question still remains: What is the purpose of the language of the Standard Model if viewed as a computer language? Isaac Asimov wrote a science fiction story that suggests one possible answer to this question. In Asimov's story a very powerful computer is developed and expanded until it meets all the needs of people and can answer practically any question. One day a child asks its parent "What is the purpose of the universe?" The parent says we will ask the all-knowing computer. The computer replies "Insufficient data but I will work on it." The computer grows and grows until space is not big enough for it so it moves itself to SuperSpace and eventually fills it up as well. Eventually long after the race of its creators has disappeared and it is the only thing remaining in existence the computer finds the answer and says "Let there be Light."

Is Asimov's story an answer? Does the language of the Standard Model relate to the fundamental purpose of the universe? Is the universe a computation? For Whom?

# 8

# A SuperString Quantum Turing Machine

**Introduction**
In this chapter we will see how SuperString Theory stacks up as the successor to the Standard Model. In an earlier chapter we explored the idea that the Standard Model has a representation as a language with a grammar specified by the Lagrangian of the theory and with the particles playing the role of the alphabet.

The questions that we now address are "Is there a better formulation of the theory of elementary particles than the Standard Model?" and "Are particles themselves reducible to something simpler?" without reference to the formulation in the previous chapter.

**Is SuperString Theory Simpler than the Standard Model?**
SuperString theorists say the Standard Model can't be right because there are too many "so-called" fundamental particles – at least 37 particles – and because it doesn't include the force of gravity. Standard Model theorists can look at SuperString theory and say,

"Well you have 37 dimensions (degrees of freedom) to play around with in SuperString theory – a 26-dimensional space, a 10-dimensional space and a 1-dimensional space ("along the string") in the currently most acceptable SuperString theory built around heterotic strings. *Thirty-seven dimensions versus thirty-seven particles (interesting coincidence).* You create particles from space-time vibrations. We assume particles are fundamental. Aren't 37-dimensions just as bothersome as 37+ fundamental particles (counting the graviton)? Too many dimensions! You have transformed the multitude of particles into a multitude of dimensions. It remains to be seen whether SuperString theory is actually a deeper theory, or perhaps the Ptolemaic theory of particles.

The Standard Model is experimentally verified. There is no convincing evidence for SuperStrings or for the hidden collapsed dimensions of SuperString theory."

The SuperString theorist replies

"We have reduced matter to artifacts of space-time and built theories that unite gravitation with the other forces of nature in a consistent way. We are within reach of a grand unified theory of all natural phenomena based on space-time geometry – the Holy Grail of Physics. The extra dimensions we introduced may number 37 – we really count them as 26 with 10 other dimensions mapped to 10 of the 26 dimensions (although we can't explain how or why as yet). But the rotational symmetry of this 26-dimensional space closely links the dimensions together so we don't have a lot of arbitrary constants in our theories unlike the Standard Model which has many arbitrary constants in its formulation."

The Standard Model theorist responds

"If you figure out how the dimensions collapse to the four space-time dimensions of the real world, then isn't it possible that the collapse may involve the introduction of arbitrary or incalculable constants as we see in the Standard Model? Arbitrary constants can appear in a spontaneous collapse such as you envision for the collapse of the dimensions."

The SuperString theorist lapses into poetry and has the last word:

> "We will continue to strive and seek and not to yield until we find the correct formulation of the SuperString theory of the universe."

**The Ideas of SuperString Theory**
The best current attempt to develop a deeper theory than the Standard Model is called SuperString Theory. This theory attempts to explain the nature of elementary particles (What is a letter?) and how elementary particles interact with each other (What is the grammar?) based on a deeper picture consisting of strings vibrating in a space with many more dimensions than the four space-time dimensions with which we are familiar.

The linguistic representation of the Standard Model described in an earlier chapter is an interesting preliminary to the SuperString Theory. We will look at SuperString Theory from a Quantum Grammar™ and Quantum Turing Machine perspective in this chapter.

There is no direct connection between the *strings* of particles (which are actually words made of characters) in our linguistic representation of the Standard Model and the SuperStrings described in this chapter. The one thing these approaches have in common is that they are both one-dimensional. A string of characters is one-dimensional – the characters are lined up in a row. A string in SuperString Theory is one-dimensional – it is a mathematical construct that is analogous to a string made of thread.

A string in the linguistic representation of the Standard Model is a discrete list of individual particle characters. A string in String Theory is a continuous, one-dimensional space (like a thread) representing one particle. It can be visualized as a rubber band (closed string representation of a particle) or as a piece of cotton thread with two ends (open string representation of a particle).

Particles in the Standard Model are point-like. SuperString Theory treats each "fundamental" particle – quarks, electrons, and so on – as an extended string in space-time. The strings are so small that their string-like nature cannot be verified experimentally with today's experiments. Current experiments see particles as point-like. If experiments could probe more closely they might see particles as strings. Strings are thought to be approximately 100 billion billion times smaller than a proton.

Some versions of String Theory assume a particle is a closed string. This type of string is often implemented in a form called a *heterotic* string. A heterotic string consists of two strands (see previous chapter for an example). One strand is a 10-dimensional string. The other strand is a 26-dimensional string. This composite string has properties that appear to be closer to reality than other types of strings.

In SuperString Theories elementary particles are vibrations of the strings. A string can vibrate in a number of different ways just like a violin string can vibrate at different frequencies. Each vibration of a SuperString corresponds to a different particle. SuperString theory should have a vibration for each different type of elementary particle including gravitons – the particle that carries the force of gravity. Actually SuperString theories predict many more particles than the particles that we have actually found in nature. Eliminating these extra particles – often by suggesting that they are too massive to be observed by current particle accelerators – is a major task of SuperString theory.

In 2500 years we have moved from the music of the spheres to the music of the strings: from divine harmony to the quantum string.

SuperString theories have many interesting features that are described in popular[14] and technical[15] books. Perhaps the most interesting feature of SuperString theories is their high degree of symmetry. First they have the rotational symmetry of a 26-dimension space-time. Then they have conformal symmetry – a symmetry that can crudely be described as stretchability – a string can be stretched or deformed in mathematical ways. They also have a peculiar symmetry called Supersymmetry. Supersymmetry is based on the rotation of fermions (spin ½ particles, spin 3/2 particles, and so on) and bosons (spin 0 particles, spin 1 particles, and so on) into each other.

Before Supersymmetry, fermions were fermions and had special properties that distinguished them. For example, they had Fermi-Dirac statistics based on the Pauli Exclusion Principle that stated that no two fermions could have exactly the same set of quantum numbers.

Before Supersymmetry, bosons were bosons and they had their own special features. Bosons had Bose statistics that stated that any number of bosons could have the same set of quantum numbers. These differences in statistics and other features gave bosons and fermions radically different properties.
Supersymmetry enables rotations to take place in a mathematical space of bosons and fermions. Fermions and bosons can be rotated into each other. As a result of Supersymmetry fermions and bosons are interrelated and form families. This symmetry puts important restrictions on theories that support it.

At the moment no solid experimental evidence exists that supports the idea that Supersymmetry is a symmetry of Nature. However the Supersymmetry concept is very appealing theoretically and it is an integral part of SuperString theories.

Having seen the basic ideas of SuperString theories we now try to relate them to the linguistic representation of the Standard Model. Is SuperString theory interpretable as a computer language?

**Equivalence of Standard Model & SuperString Theory?**
The particle representation of the Standard Model is one theoretical description of nature. The SuperString theory may be an alternate description of nature based on vibrating strings in a higher-dimensional space-time.

There are simple solvable models that have been discovered in recent years that suggest that these theories could eventually grow into complementary representations. These simple models have a space-time formulation that contains miniature black holes and no matter. This formulation can be transformed into another representation that transforms the black holes into what appear to be particles. There is a geometrical space-time formulation without matter – and an equivalent particle formulation.

These equivalent formulations may be hints of an equivalence between a suitably extended Standard Model of particles and a SuperString theory.

If we think of computer languages that are equivalent to each other but differ greatly in their appearance, can we view the Standard Model (*suitably extended*) and SuperString theory as potentially equivalent formulations? Particles are the fundamental symbols in the Standard Model Quantum Grammar. Perhaps space-time geometrical objects are the fundamental symbols in the Quantum Grammar of SuperString Theory.

**Linguistic Model of SuperString Theory**
If we take the idea of a computer language description of Nature seriously, then we have to understand how the features of Supersymmetry theory can be viewed as symbols of a computer language grammar and what the grammar of this language is.

---
[14] Michio Kaku, *Hyperspace*, (Bantam, Doubleday, Dell Publishing, New York, 1994) among others.
[15] Joseph Polchinski, *String Theory* (Cambridge University Press, New York, 1998). D. Bailin and A. Love, *Supersymmetric Gauge Field Theory and String Theory* (Institute of Physics Publishing, Philadelphia, PA, 1994) among others.

We can view the Standard Model as a computer language and the Supersymmetry theory as a possibly more fundamental (lower level) computer language. Is the Supersymmetry language the machine language upon which the Standard Model language is based? In this regard one thinks of computer languages such as C++ that are translatable to a lower level language such as assembly language and then to an even lower level language machine language. The hallmark of these lower level languages is that they are increasingly simpler in nature with fewer and fewer features.

The lowest level computer language, machine language, has few constructs in it and directly tells the computer chips what to do. On the other hand, the interpretation of the constructs of a higher level language in terms of lower level language constructs is often complex.

**A Language and Turing Machine for SuperString Theory**

The natural choice for a grammar for a SuperString Theory has letters (terminal symbols) that are SuperStrings. Each letter represents a specific string vibration. The production rules for the SuperString grammar are simple at the symbolic language level. The details of the numeric calculations behind the grammar rules are complex. They are described in books such as those listed in reference 26.

A set of grammar rules for the linguistic representation of SuperString interactions is:

Rule 1.    ◎  →  ◎ ◎

Rule 2.    ◎◎  →  ◎

Rule 3.    ◎◎  →  ◎′ ◎′

The quote marks on the right side of rule 3 indicate that the symbols on the right side represent different particles (terminal symbols) from the particles on the left side. This notation is part of the formalism of computer language grammar rules. The grammar rules can be represented pictorially as closed string diagrams:

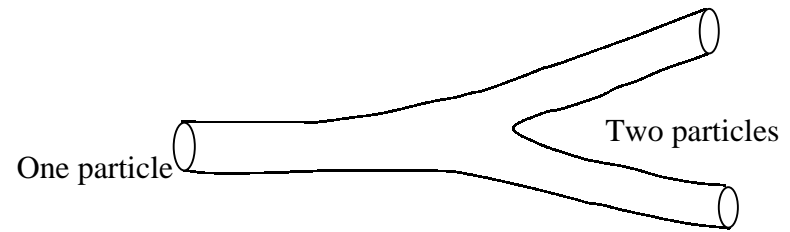

Figure. Two visualizations of rule 1 for the "fission" of a string into two strings.

The "fission" of a photon string into an electron string and a positron string is an example of this grammar rule. The strings look like tubes or pipes as they move in time: if you take many pictures of a moving string and superimpose the pictures the motion of the strings will trace out tubes. That is why the visualizations appear as tubes.

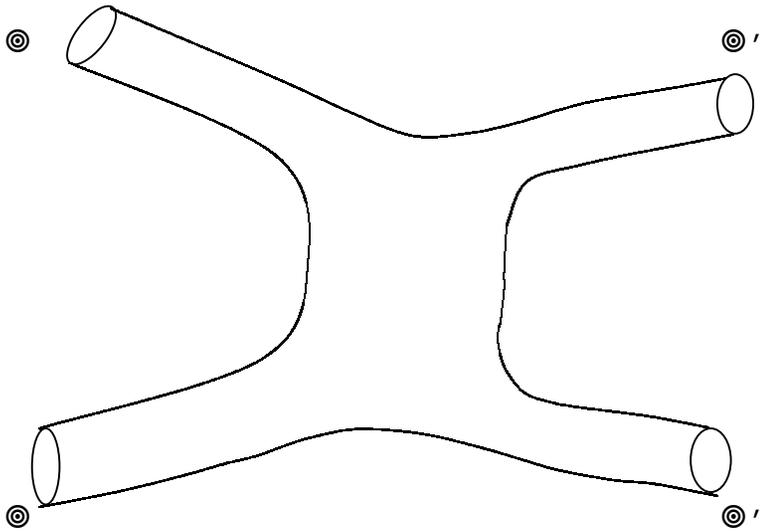

Figure. A diagrammatic visualization of rule 3. The strings look like tubes or pipes as they move in time. In this case two strings interact and then produce two output strings.

Notice the grammar production rules for SuperStrings are much simpler than the grammar production rules for the Standard Model. This was to be expected if the SuperString Theory is in some sense a deeper theory corresponding to a lower level computer language – perhaps the equivalent of machine language for particles.

The Quantum Computer that corresponds to SuperString theory accepts input states consisting of a word containing a string of particle strings. It processes the input and generates an output word consisting of a symbol string consisting of particle strings. Each possible output word has a corresponding probability of being produced since the Quantum Computer is probabilistic. Inside the Quantum Computer the input particles interact – producing a variety of intermediate particle states or words. Eventually the output particles are generated.

The general features of the SuperString Quantum Turing Machine (SQTM) are similar to the Quantum Turing Machine for the Standard Model. However the language and the theory behind the language are different. The SQTM generates intermediate string states based on the details of the SuperString theory that it implements. An example of an interaction between strings with interesting intermediate states is:

◎◎ → ◎′◎′◎ → ◎′◎′◎′◎′ → ◎′◎′◎′ → ◎′ ◎′

The intermediate strings are generated using the three SuperString grammar rules stated earlier. This set of transitions can also be depicted with a SuperString theory diagram with time orderings:

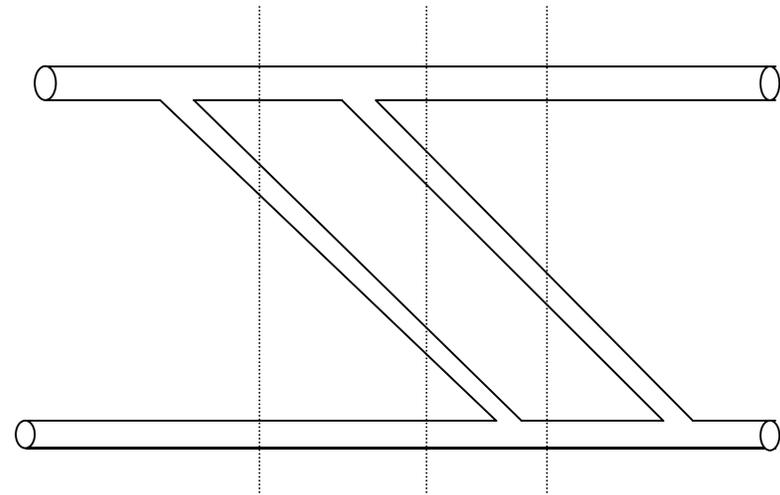

Figure. String diagram of two interacting strings that generate two strings in intermediate states. These generated strings are absorbed so that only two strings emerge in the output. The intermediate state "words" are indicated by vertical dashed lines. Many other diagrams contribute to this string-string interaction when realistic calculations are made.

The SQTM offers a different view of the features of SuperString Theory. This view may be helpful in deepening our understanding of the implications of SuperString Theory and its relation to the Standard Model.

## Beyond SuperStrings – Hidden Dimensions as Quantum Computer Tapes

SuperString theories assume the existence of hidden dimensions that are the origin of the internal symmetries of elementary particles. These dimensions are somehow compactified (curled up) in a manner that is not well understood. As a result we only experience the normal four dimensions of space and time.

The Quantum Computer representations of the Standard Model and Supersymmetry that we have developed suggest an alternate possibility: the internal tapes of the Quantum Computer may define a different space from the space of the input and output particles.

A Quantum Computer has several tapes. These tapes play the role of computer memory just like computer tapes. One tape – the input tape – has the input particles on it. Another tape – the output tape – has the output particles on it. (The input and output tapes could be the same tape.)

The Quantum Computer also has an internal tape that can be viewed as a set of tapes or as a multidimensional tape upon which intermediate states of the particles can be stored. We can view a Quantum Computer that processes an input state to produce an output state as:

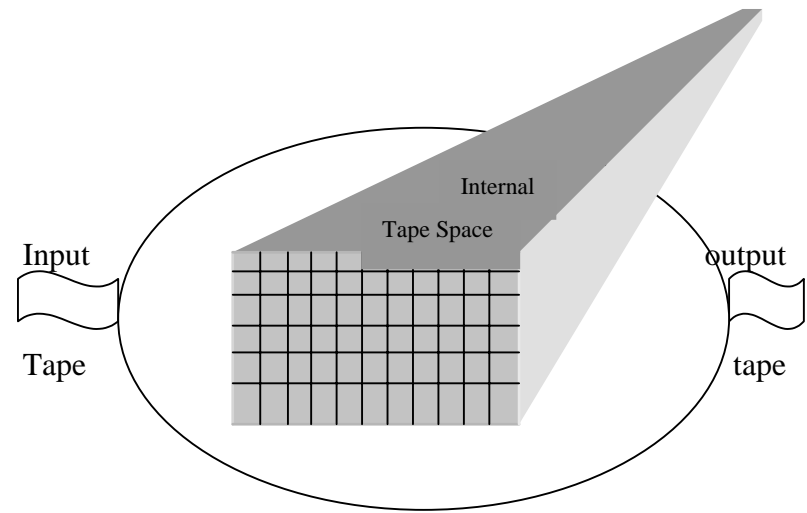

Figure. A Quantum Computer with an internal multidimensional space (tape).

The internal Turing machine tape defines an internal "space" to which we can assign dimensions. This space, as it is normally viewed in Turing machines, is divided into cells. However in Quantum Computers it can be made into a continuous space. Further we can make this space the space of the internal symmetries of the Standard Model or Supersymmetry Theory (or in part the space of symmetries of the Standard Model or Supersymmetry Theory).

This concept leads us to view the space of internal symmetries as part of the Quantum Computer. This space is relevant when particles interact (collide). The internal space is not relevant and is not "visible" when particles are not interacting. This simple view agrees with our experience. The "space" of the input and output tapes interfaces with (or melds into) the internal space when particles interact. So we can expect that the internal space somehow meshes the input and output space in a smooth way.

The smoothness (continuity) of the transition between the internal space of the Quantum Computer and the external space of the input and output states could result in a unified theory that unites gravity and the internal particle symmetries. This approach is similar in flavor to the standard SuperString approach for the unification of the forces of nature.

An interesting point related to the Quantum Computer representation of the Standard Model and Supersymmetry is quark confinement. Individual, isolated quarks have not been found in Nature. Quarks always appear bound together with other quarks: three quarks in a proton, a quark-antiquark pair in a pion, and so on. The inability to split a proton or any other particle containing quarks into individual quarks is an interesting phenomenon. Most theorists believe it is a feature of the Strong Interaction between quarks. Unlike other forces such as gravity the Strong Interaction becomes stronger and stronger as quarks move further apart preventing them from separating. This feature of the Strong Interaction can be restated in the grammar of the Standard Model or Supersymmetry by saying that quarks are *non-terminal symbols*.

A non-terminal symbol is a symbol that can appear within the grammar of a language but cannot be an input symbol or an output symbol. It only appears as an intermediate symbol within the grammar rules. Protons, pions and other composite particles composed of quarks can be viewed as *terminal symbols* – symbols that can appear as input symbols or output symbols. Quarks only appear inside the Quantum Computer as particles in intermediate states generated by the grammar. They are reassembled into composite particles that are output (terminal) symbols.

The Quantum Computer representations of the Standard Model and Supersymmetry provide an alternate view of elementary particles that is both simple and yet capable of embodying significant features of these theories in a straightforward way. Quark confinement is one Standard Model feature that is easily represented in the linguistic representation. The existence of hidden dimensions is also easily represented in the Quantum Computer approach as an internal multidimensional tape. The linguistic approach to elementary particle theories naturally accommodates major features of elementary particles.

# 9

# Quantum Computers as the Foundation of SuperStrings

**Introduction**
Many physicists feel that SuperString Theory is the next step beyond the Standard Model. This theory is based on an interesting idea: *build matter out of the fabric of space and time*. Not the space and time with which we are familiar but rather a space and time of many more dimensions. Most of the dimensions of this higher dimensional universe curl up into a "tight ball" so we only see the four dimensions of space and time of everyday experience. The curled extra dimensions are thought to be the origin of the internal symmetries that particles possess in the Standard Model.

The reduction of matter to a feature of space-time is an important concept. It leads us to *an insubstantial universe consisting only of space and time that is based on quantum probabilities*. It lacks the solidity of our world of ordinary perceptions. A world governed by a language and consisting of nothingness. Matter being mere wrinkles in space and time.

Actually space and time themselves – since they are represented by the gravitational field which itself is a SuperString construct – also become constructs. In the end we wind up with "nothing" – a "nothing" that develops structure through the SuperString Theory. The structure "becomes real" through the existence of SuperStrings – particles represented by mathematical entities that resemble open and closed strings. "Normal" space and time now exist through the presence of gravitons – particles carrying the gravitational force – which are in reality mathematical strings.

In the previous chapter we saw the appearance of both Fermi and Bose fields in Quantum Computers. We also saw the apparent similarity of the lattice space of the Quantum Computer memory to a space that could support Supersymmetry. These observations lead us to consider the possibility that a Quantum Computer formalism may provide a foundation for SuperString Theory.

SuperString theory today has some notable successes. However it is still a patchwork quilt of clever ideas that lead to something approaching a realistic theory of elementary particles. The situation is reminiscent of the Bohr model of the atom. It blended together disparate ideas (that could be viewed as incompatible in their origin) to produce a model of the Hydrogen atom that explained some major experimental features of Hydrogen. It was not until Quantum Mechanics developed about ten years later that a logically satisfactory theory of Hydrogen was created.

SuperString theory may be a "Bohr model" of elementary particles in wait for the right theory. The SuperString Quantum Computer that will be developed in this chapter offers a totally new view of the origins of SuperString theory.

## The Bosonic String Part of a Quantum Computer

The beginning point is the standard view of a Quantum Computer augmented to contain a set of tapes for memory and a set of linear one-dimensional arrays of tape heads. Each array has an infinite number of tape heads and looks like a "tape" (or string) of tape heads.

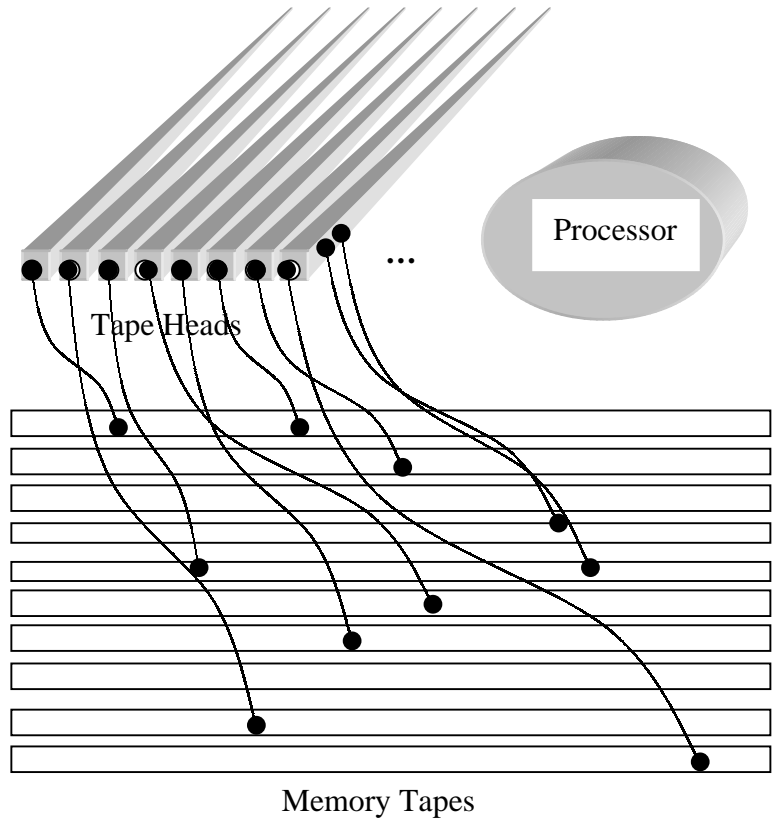

Figure. A Polycephalic Quantum Computer™ with infinite arrays of tape heads and many memory tapes.

We will call an infinite linear array of tape heads a *string of tape heads*.

Following a similar path to the discussion in the preceding chapter we associate Fermi creation and annihilation operators with each tape location (or 2-state observable), and Bose creation and annihilation operators with each tape head.

In the preceding chapter we defined a string of tape heads with corresponding operators $c_n$ with n ranging from $-\infty$ to $\infty$. We can redefine these operators[16] to show they can eventually be used to develop a SuperString formalism. Let us define

$$a_n = c_n \qquad n \geq 0$$
$$a_n^\dagger = c_n^\dagger \qquad n \geq 0$$

$$\underline{a}_{-n} = c_n \qquad n < 0$$
$$\underline{a}_{-n}^\dagger = c_n^\dagger \qquad n < 0$$

The operators $a_n$ and $\underline{a}_n$ and their hermitean conjugates can be generalized to multiple strings of tape heads by adding an index specifying the tape head string.

Bose operators for D strings of tape heads are:

$$a_n^\mu \text{ and } \underline{a}_n^\mu$$

with n ranging from 0 to $\infty$. The index $\mu$ specifies the tape head string and ranges from 0 to D.

---

[16] We attempt to use a notation consistent with standard texts on SuperStrings: D. Bailin and A. Love, *Supersymmetric Gauge Field Theory and String Theory* (Institute of Physics Publishing, Philadelphia, PA, 1994) page 157; Joseph Polchinski, *String Theory* (Cambridge University Press, New York, 1998)

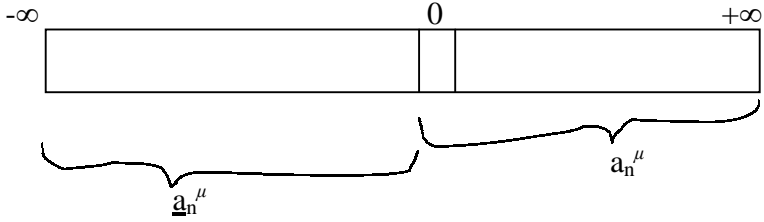

Figure. The $\mu^{th}$ string of tape heads. There is a tape head on the string for each value of n with corresponding operators $a_n^\mu$ for $n \geq 0$ and $\underline{a}_{-n}^\mu$ for $n < 0$.

These operators have the same tape head commutation relations seen in the previous chapter. Taking account of the added index the commutation relations become:

$$[a_i^\mu, a_j^{\nu\dagger}] = -\delta_{ij}\eta^{\mu\nu}$$

$$[a_i^\mu, a_j^\nu] = 0$$

$$[a_i^{\mu\dagger}, a_j^{\nu\dagger}] = 0$$

where $\eta^{\mu\nu}$ is non-zero if $\mu \neq \nu$ and has absolute value 1 if $\mu = \nu$. The tape head operators for different tape head strings commute with each other. The operators $\underline{a}_i^\mu$ and $\underline{a}_j^{\nu\dagger}$ have similar commutation relations.

To establish a connection with the SuperString literature and to make a connection with the bosonic strings of Physics we introduce the operators $\alpha_n^\mu$ and $\underline{\alpha}_n^\mu$ with n ranging from $-\infty$ to $\infty$. The index $\mu$ again specifies the tape head string. These operators are related to $a_n^\mu$ and $\underline{a}_n^\mu$ via:

$$\alpha_n^\mu = \sqrt{n}\, a_n^\mu \qquad \alpha_{-n}^\mu = \sqrt{n}\, a_n^{\mu\dagger} \qquad \text{for } n > 0$$

and

$$\underline{\alpha}_n^\mu = \sqrt{n}\, \underline{a}_n^\mu \qquad \underline{\alpha}_{-n}^\mu = \sqrt{n}\, \underline{a}_n^{\mu\dagger} \qquad \text{for } n > 0$$

These operators are the oscillator coefficient operators in the mode expansion of the closed bosonic string:

$$X^\mu = x^\mu + l^2 p^\mu \tau + .5l \sum_{n \neq 0} (\alpha_n^\mu e^{-2in(\tau - \sigma)} + \underline{\alpha}_n^\mu e^{-2in(\tau + \sigma)})/n$$

Notice we have effectively created a field just as we did in the previous chapter – although with the complications of an extra index. The coordinate $X^\mu$ is a coordinate in a D dimensional space. The parameters $\tau$ and $\sigma$ can be viewed as parametrizing a closed string in D dimensional space with a time-like coordinate $\tau$ and a space-like coordinate $\sigma$. For fixed $\tau$, $X^\mu$ traces out a curve (a string) as $\sigma$ varies. The terms in the summation are for a "right mover" part that depends on $\tau - \sigma$ and uses the operators $\alpha_n^\mu$, and for a "left mover" part that depends on $\tau + \sigma$ and uses the operators $\underline{\alpha}_n^\mu$. The left mover modes are used to construct heterotic strings – the type of string that appears to result in a theory that is closest to reality.

The mode expansion of $X^\mu$ can be viewed as the solution of the one-dimensional wave equation:

$$\left(\frac{\partial^2}{\partial \tau^2} - \frac{\partial^2}{\partial \sigma^2}\right) X^\mu = 0$$

Introducing a metric $h_{\alpha\beta}(\tau, \sigma)$ in addition to the metric $\eta^{\mu\nu} = \text{diag}(1,-1)$ that appeared earlier in the commutation relations we can develop the mode expansion for $X^\mu$ from the action for the relativistic bosonic string

$$S = \frac{-T}{2}\int d\tau \int d\sigma\, (-\det h)^{1/2}\, h^{\alpha\beta}\eta^{\mu\nu}\partial_\alpha X_\mu \partial_\beta X_\nu$$

We do this by obtaining the Euler-Lagrange equations of motion, and using reparametrization invariance and conformal invariance to reduce the metric $h_{\alpha\beta}$ to

$$h_{\alpha\beta} = \eta_{\alpha\beta}$$

which then leads to the above wave equation.

The other features of bosonic strings (such as the Virasoro algebra) can be obtained by consistently applying the standard field theoretic formalism (reference 26).

As a result of this development we can see the tape heads of a Quantum Computer supplemented by reparametrization and conformal invariance can be seen as providing a representation of the bosonic string concept. Consistency requirements select the dimension D of the bosonic string space to be 26. Thus there are 26 infinite "strings" of tape heads in the Quantum Computer for SuperString theory.

This result provides a new concrete representation of bosonic strings. In the next section we will see the memory tapes of a Quantum Computer can be mapped to SuperStrings.

**The SuperString Part of a Quantum Computer**
In the preceding section we saw how arrays (strings) of tape heads could be related to bosonic strings. In this section we show the Fermi operators of memory tapes can be related to SuperStrings. In the next section we join the bosonic string discussion together with the results of this section to show how heterotic strings can naturally emerge from the parts of a Quantum Computer.

The preceding chapter showed that each tape position could be associated with creation and annihilation operators $b_i$ and $b_j^\dagger$ satisfying the anticommutation relations:

$$\{ b_i, b_j^\dagger \} = \delta_{ij}$$

$$\{ b_i, b_j \} = 0$$

$$\{ b_i^\dagger, b_j^\dagger \} = 0$$

where $\delta_{ij}$ equals 1 if $i = j$ and $\delta_{ij}$ equals zero otherwise. The indices i and j are integers that label tape positions and range between $-\infty$ and $+\infty$.

Let us now consider the Polycephalic Quantum Computer™ pictured at the beginning of this chapter that has a number of memory tapes. Let us suppose there are D tapes that are labeled with the index $\mu$.

In order to make our discussion similar in form to the Physics literature (reference 26 and references therein) we now define operators for Ramond boundary conditions:

$$\begin{aligned} d_n &= b_n & n \geq 0 \\ d_{-n} &= b_n^\dagger & n \geq 0 \end{aligned}$$

$$\begin{aligned} \underline{d}_{-n} &= b_{n-1} & n \leq 0 \\ \underline{d}_{-n} &= b_{n-1}^\dagger & n \leq 0 \end{aligned}$$

These new operators satisfy

$$\{ d_i, d_j \} = \delta_{i+j,0}$$
$$\{ \underline{d}_i, \underline{d}_j \} = \delta_{i+j,0}$$

It is easy to generalize the above anticommutation relations to:

$$\{ d_i^\mu, d_j^\nu \} = -\delta_{i+j,0} \eta^{\mu\nu}$$
$$\{ \underline{d}_i^\mu, \underline{d}_j^\nu \} = -\delta_{i+j,0} \eta^{\mu\nu}$$

where i and j range from $-\infty$ to $\infty$.

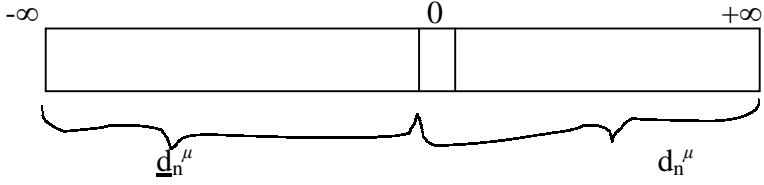

Figure. The $\mu^{th}$ memory tape string. There is a location on the string for each value of n with corresponding operators $d_n^\mu$ for $n \geq 0$ and $\underline{d}_{-n}^\mu$ for $n < 0$.

We can now form the mode expansions for a fermionic *right mover*, closed SuperString for periodic (Ramond) boundary conditions (reference 26):

$$\Psi_R^\mu = \sum_{n \in Z} d_n^\mu \, e^{-2in(\tau - \sigma)}$$

and for anti-periodic (Neveu-Schwartz) boundary conditions:

$$\Psi_R^\mu = \sum_{n \in Z + 1\backslash 2} b_n^\mu \, e^{-2in(\tau - \sigma)}$$

where the sum extends from $-\infty$ to $\infty$ and where the operators $b_m^\mu$ are the corresponding Neveu-Schwartz operators with m half-integral.

The operators $b_m^\mu$ are not the operators $b_n$ defined on the previous page. The operators $b_n^\mu$ and $\underline{b}_n^\mu$ satisfy similar anticommutation relations to $d_n^\mu$ and $\underline{d}_n^\mu$ (see below). They are an alternate set of creation and annihilation operators for the memory tapes labeled with half integer values.

We can now form the mode expansions for a fermionic *left mover*, closed SuperString assuming periodic (Ramond) boundary conditions:

$$\Psi_L^\mu = \sum_{n \in Z} \underline{d}_n^\mu \, e^{-2in(\tau + \sigma)}$$

and anti-periodic (Neveu-Schwartz) boundary conditions:

$$\Psi_L^\mu = \sum_{n \in Z + 1\backslash 2} \underline{b}_n^\mu \, e^{-2in(\tau + \sigma)}$$

where the sum extends from $-\infty$ to $\infty$. These mode expansions satisfy the equations:

$$\left(\frac{\partial}{\partial \tau} + \frac{\partial}{\partial \sigma}\right) \Psi_R^\mu = 0$$

$$\left(\frac{\partial}{\partial \tau} - \frac{\partial}{\partial \sigma}\right) \Psi_L^\mu = 0$$

and the canonical anti-commutation relations:

$$\{\Psi_R^\mu(\tau, \sigma), \Psi_R^\nu(\tau, \sigma')\} = \{\Psi_L^\mu(\tau, \sigma), \Psi_L^\nu(\tau, \sigma')\}$$

$$= -2\pi\delta(\sigma - \sigma')\eta^{\mu\nu}$$

$$\{\Psi_R^\mu(\tau, \sigma), \Psi_L^\nu(\tau, \sigma')\} = 0$$

If we form the spinor

$$\Psi^\mu(\tau, \sigma) = \begin{pmatrix} \Psi_R^\mu(\tau, \sigma) \\ \Psi_L^\mu(\tau, \sigma) \end{pmatrix}$$

then we can define an action that leads to the above differential equations through the canonical approach from Euler-Lagrange equations of motion in the covariant gauge. The combined action for SuperStrings (SuperSymmetry requires a bosonic string term) is

$$S_0 = \frac{-1}{2\pi} \int d^2\sigma \, (-\det h)^{\frac{1}{2}} \, (h^{\alpha\beta} \partial_\alpha X^\mu \partial_\beta X_\mu + i\Psi^\mu \rho^\alpha \partial_\alpha \Psi_\mu)$$

where

$$\rho^0 = \begin{pmatrix} 0 & -i \\ i & 0 \end{pmatrix} \qquad \rho^1 = \begin{pmatrix} 0 & i \\ i & 0 \end{pmatrix}$$

Consequently the entire formalism of SuperStrings can be "reverse engineered" from memory tapes in a Quantum Computer.

Consistency requirements select the dimension D of the SuperString space to be 10. Thus there are 10 infinite memory tapes, or memory "strings", in the Quantum Computer for SuperString theory.

The SuperString action also provides a 10-dimensional bosonic string sector. The memory of the Quantum Computer could be augmented to support an additional 10 tapes with bosonic observables (not 2-state observables) supporting a spectrum consisting of the positive integers. This addition would provide consistency with the action.

The next section joins bosonic strings and fermionic Superstrings to create a heterotic string. Heterotic string theories appear to offer the closest connection to Reality as we currently know it.

## The Quantum Computer as the Foundation of Heterotic Strings

In the preceding two sections we have shown how to formulate a Quantum Computer equivalent to Superstring theory consisting of bosonic strings (from tape heads) and fermionic SuperStrings (from memory tapes). In this section we will see how to build a heterotic string within the framework of the SuperString Quantum Computer.

The SuperString Quantum Computer has ten infinite memory tapes with fermionic operators that support a 10-dimensional SuperString space with SuperStrings. It also has twenty-six infinite arrays or strings of tape heads with boson operators that support a 26-dimensional space for bosonic strings.

A heterotic string is a closed string created by using right movers of a (type II) SuperString and the left movers of a bosonic string. The space of states of heterotic strings is the direct product of the spaces of the right and left movers.

There is an apparent problem because the space of the left movers (bosonic strings) is 26 space-time dimensions while the space of the right movers (SuperStrings) is 10 space-time dimensions. The resolution of the problem is to compactify 16 of the left mover dimensions by associating them with a 16-dimensional torus with extremely small radii of the order of fundamental string length. So 16 dimensions shrink to almost nothing leaving 10 "real" dimensions for the left mover. These extra dimensions presumably play a useful role by generating the internal symmetries of the elementary particles (as found in the Standard Model) through the Kaluza-Klein mechanism. In the Kaluza-Klein mechanism a higher dimensional theory containing gravity generates Yang-Mills-like gauge fields when the number of dimensions is reduced through compactification.

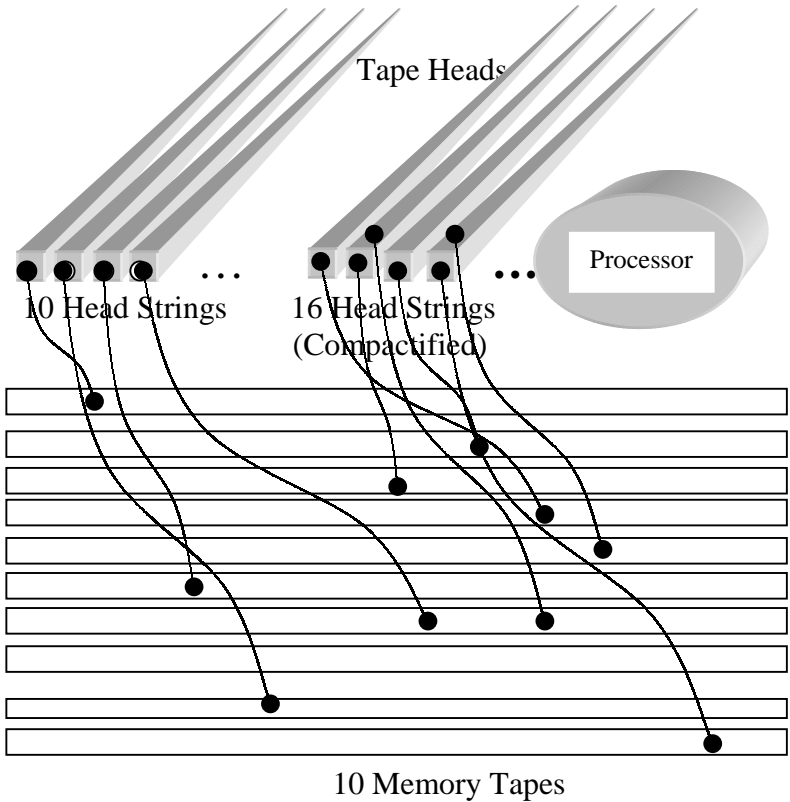

Figure. SuperString Quantum Computer with 26 infinite arrays (strings) of tape heads and 10 memory tapes.

The creation and annihilation operators of the tape heads and memory tapes can be used to create states corresponding to elementary particles. For example, the combined tape head-memory location state

$$b^i_{-\frac{1}{2}}|0\rangle_R \, \underline{a}^j_{-1}|0\rangle_L \qquad i, j = 1, \ldots 8$$

can be decomposed into a traceless symmetric 10-dimensional graviton, an anti-symmetric tensor and a scalar dilaton. The state is defined using $|0\rangle_R$ as the vacuum state (empty state) for right movers and $|0\rangle_L$ as the vacuum state (empty state) for left movers, with $b^i_{-\frac{1}{2}}|0\rangle_R$ being right mover SuperStrings of the Neveu-Schwartz type and with $\underline{a}^j_{-1}|0\rangle_L$ being bosonic left mover strings.

The combined role of the tape heads and the memory tapes supports the creation of heterotic SuperString particle states.

The next issue that the SuperString Quantum Computer must address is interactions between the SuperStrings. SuperString theory does not have an intrinsic well-defined way of specifying interactions between SuperStrings. So we cannot look for strong guidance from SuperString theory when we try to develop interactions within the framework of the SuperString Quantum Computer.

SuperString Theory does have a principle that is used to specify interactions – an isomorphism between states and vertex operators. This principle is a feature of Conformal Field Theories. SuperString theories are Conformal Field Theories.

The vertex operators of SuperString theories are used to define interactions between SuperStrings. A vertex operator is necessary whenever SuperStrings "intersect" – when a SuperString fissions into two SuperStrings or two SuperStrings amalgamate to form one SuperString.

Since we know how to define SuperString states using the raising (creation) and lowering (annihilation) operators of the SuperString Quantum Computer we can use that information and the isomorphism (correspondence) to define vertex operators within the SuperString Quantum Computer. In fact we can define states corresponding to a vertex operator in the SuperString Quantum Computer processor.

In view of its role the SuperString Quantum Computer processor appears to be the natural place to specify SuperString interactions. The processor is supposed to contain the program of the computer and the initial state of the computer – the input.

An alternative to using the processor to specify interactions is to use a standard vertex operator expressed in SuperString raising and lowering operators. An example is the normal ordered vertex operator:

$$:e^{ik \cdot X}:$$

where $k^\mu$ is the momentum of a SuperString particle and $X^\mu$ is the SuperString coordinate operator.

## Is the SuperString Quantum Computer More Fundamental than Coordinate Space SuperString Theory?

The view that we have espoused in this discussion is that the foundation of SuperStrings is in Quantum Computers. The raising and lowering operators are the key features of the formalism. If one looks at the development of SuperString theories (reference 15 and references therein) then it is very evident that the detailed study of almost all SuperString features is based on raising (creation) and lowering (annihilation) operator expressions.

In the standard development of SuperString Theory raising and lowering operators are "derived" from a space-time formalism filled with nice symmetries. In our approach the raising and lowering operators of the SuperString Quantum Computer are the fundamental constructs and the space-time formalism is secondary.

SuperString Quantum Computer™ ⟹ Space-Time Theory

Establishing a Quantum Computer foundation for SuperString Theory changes the perspective from a space-time formalism to a more fundamental computational formalism. We have freed SuperStrings from the trappings of space-time!

One might argue that space-time approaches to the development of physical theories have been the only successful approach in the past. But one might also argue that a space-time approach is meaningless at ultra-short distances where space and time itself are in question. The Quantum Computer formulation avoids this issue by basing SuperString theory on computation.

On the other hand, one could argue that a space-time approach should not be ruled out just because the distances are ultra-short. It works well at the larger distances with which we are familiar. Perhaps viewing short distance physics as different because we can't "see" at that level is the subtle head of anthropomorphism rearing up. Why shouldn't the space-time approach work equally as well at the short distances that we cannot directly perceive as the larger distances that we can directly perceive?

Well there is one example where the extrapolation from common experience to shorter distances has abjectly failed: quantum phenomena. The theories of mechanics and electromagnetism worked beautifully at the level of everyday experience. We needed a major change – quantization - to develop a theory that worked at the sub-atomic short distance scale. Is the foundation of SuperString theory a new level with Computational Quantization?

# 10

# Quantum Computer Processor Operations and Quantum Computer Languages

## Introduction

A natural question that arises when one considers Quantum Computers is the role of the Quantum Computer processor and the operations it supports. A further question of some interest is whether a quantum machine language exists and what its nature might be. Lastly the question of higher level languages is also relevant. Can we develop a Quantum Assembly Language™? What is the nature of High Level Quantum Languages™? Are there, for example, equivalents to the C or C++ languages?

## Computer Machine and Assembly Languages

The traditional (non-quantum) computer can be viewed simply as a main memory, an accumulator or register (modern computers have many registers), and a central processing unit (CPU) that executes a program (instructions) step by step. It can be visualized as:

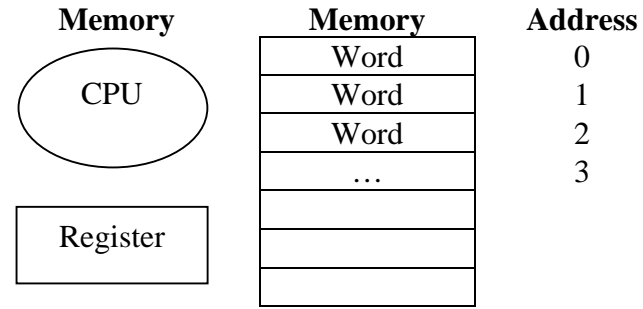

Figure. Simplified model of a normal computer.

A set of data and a program (or set of instructions) is stored in memory and the CPU executes the program step by step using the data to produce an output set of data.

The basic instructions of assembly language and machine language move data values between memory and the register (or registers), manipulate the data value in the register and provide basic arithmetic and logical operations[17]:

    LOAD M  –  load the value at memory location M into the register

    STORE M  –  store the value in the register at memory location M

    SHIFT k  –  shift the value in the register by k bits

---

[17] See for example Kurt Maly and Allen R. Hanson, *Fundamentals of the Computing Sciences* (Prentice-Hall, Inc., Englewood Cliffs, NJ, 1978) Chap. 8.

The following arithmetic instructions modify the value in the register. The AND, OR and NOT instructions perform bit-wise and, or and not operations.

    ADD M     – add the value at memory location M to the value in the register

    SUBTRACT M   – subtract the value at memory location M from the value in the register

    MULTIPLY M   – multiply the value in the register by the value at memory location M

    DIVIDE M   – divide the value in the register by the value at memory location M

    AND M –     change the value in the register by "anding" it with the value at memory location M

    OR M –     change the value in the register by "oring" it with the value at memory location M

    NOT –     change the value in the register by "not-ing" it

The following instructions implement input and output of data values.

    INPUT M  – input a value storing it at memory location M

    OUTPUT M  – output the value at memory location M

A computer has another register called the Program Counter. The value in the program counter is the memory location of the next instruction to execute. The following instructions support non-sequential flow of control in a program. A program can "leap" from one instruction in a program to another instruction many steps away and resume normal sequential execution of instructions.

    TRA M   – set the value of the program counter to the value at memory location M

    TZR M   – set the value of the program counter to the value at memory location M if the value in the register is zero.

    HALT    – stop execution of the program

The above set of instructions form an extremely simple assembly language. They also are in a one-to-one correspondence with machine instructions (machine language). Most assembly and machine languages have a much more extensive set of instructions.

## Algebraic Representation of Assembly Languages

The normal view of assembly language is that it has a word or instruction oriented format. Some assembly language programmers would even say that assembly language is somewhat English-like in part.

Computer languages in general have tended to become more English-like in recent years in an attempt to make them easier for programmers. Some view a form of highly structured English to be a goal for computer programming languages.

In this section we follow the opposite course and show that computer languages can be reduced to an algebraic representation. By algebraic we mean that the computer language can be represented with operator expressions using operators that have an algebra similar to that of the raising and lowering operators seen earlier. We will develop the algebraic representation for the case of the simple assembly language described in the previous section. There are a number of reasons why this reduction is interesting:

1. It may help to understand SuperString dynamics more deeply (later in this chapter).

2. It will deepen our understanding of computer languages.

3. It provides a basis for the understanding of Quantum Computers.

4. It may have a role in research on one of the major questions of computer science: proving a program actually does what it is designed to do. Algebraic formalisms are generally easier to prove theorems then English-like formalisms.

The algebraic representation can be defined at the level of individual bits based on anti-commuting Fermi operators. But it seems more appropriate to develop a representation for "words" consisting of some number of bits. An algebraic representation for a word-based assembly language can be developed using commuting harmonic oscillator-like raising and lowering operators.

A word consists of a number of bits. In currently popular computers the word size is 32 bits (32-bit computer). The size of the word determines the largest and smallest integer that can be stored in the word. The largest integer that can be stored in a 32-bit word is 4,294,967,294 and the smallest integer that can be stored in a 32-bit word is 0 if we treat words as holding unsigned integers.

To develop a simple algebraic representation of assembly language we will assume the size of a word is so large that it can be viewed as infinite to a good approximation. (It is also possible to develop algebraic representations for finite word sizes.) As a result memory locations can contain non-negative integers of arbitrarily large value.

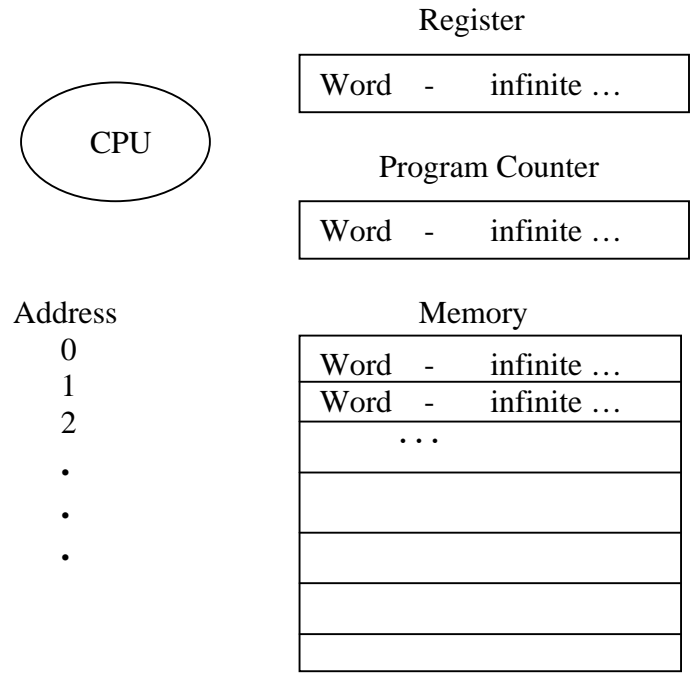

Figure. Visualization of a Computer with infinite words.

To establish the algebraic representation we associate a harmonic raising operator $a_i^\dagger$ and a lowering operator $a_i$ with each memory location. These operators satisfy the commutation relations:

$$[\, a_i, a_j^\dagger \,] = \delta_{ij}$$
$$[\, a_i, a_j \,] = 0$$
$$[\, a_i^\dagger, a_j^\dagger \,] = 0$$

where $\delta_{ij}$ is 1 if i = j and zero otherwise. We define a pair of raising and lowering operators for the register r and $r^\dagger$ with commutation relations

$$[\, r_i, r_j^\dagger \,] = \delta_{ij}$$
$$[\, r_i, r_j \,] = 0$$

$$[r_i^\dagger, r_j^\dagger] = 0$$

The ground state of the computer is the state with the values at all memory locations set to zero. It is represented by the vector

$$|0, 0, 0, \ldots\rangle \equiv |0\rangle \equiv \Phi_V$$

A state of the computer will be represented by a vector of the form

$$|n, m, p, \ldots\rangle = N (r^\dagger)^n (a_0^\dagger)^m (a_1^\dagger)^p \ldots |0\rangle$$

where N is a normalization constant and with the first number being the value in the register, the second number the value at memory location 0, the third number the value at memory location 1, and so on. For simplicity we will not consider superpositions of computer states at this point. We will discuss superpositions later in this chapter. Within this limitation we can set a computer state to have certain initial values in memory and then have it evolve by executing a "program" to a final computer state with a different set of computer values in memory. The "program" is a mapping of the instructions of an assembly language program to algebraic expressions in the raising and lowering operators.

## Basic Operators of the Algebraic Representation

The key operators that are required for the algebraic representation are:

Fetch the Value at a Memory Location (Number Operator)

$$N_m = a_m^\dagger a_m$$

For example,

$$N_m | \ldots, n, \ldots \rangle = n | \ldots, n, \ldots \rangle$$

↑

$m^{th}$ memory location value

Set the Value at Memory Location m to Zero

$$M_m = \frac{(a_m)^{N_m}}{\sqrt{N_m!}}$$

The above expression for $M_m$ is symbolic. The expression represents the following expression in which the operators are carefully ordered to avoid complications (c-numbers etc.) resulting from reordering.

$$M_m \equiv \sum_q \frac{(\ln a_m)^q N_m^q}{q!} \frac{1}{\sqrt{N_m!}}$$

where the sum ranges from 0 to ∞. When $M_m$ is applied to a state it sets the value of the $m^{th}$ memory location to zero.

$$M_m | \ldots, n, \ldots \rangle = \frac{(a_m)^n}{\sqrt{n!}} | \ldots, n, \ldots \rangle$$

↑

$m^{th}$ memory location value

$$= | \ldots, 0, \ldots \rangle$$

The repeated application of factors of $N_m$ to the state results in factors of n.

Change the Value at Memory Location m from 0 to the Value at Location n

$$P_m^{\,n} = \frac{(a_m^\dagger)^{N_n}}{\sqrt{N_n!}}$$

The above expression for $P_m^{\,n}$ is also symbolic. The expression represents the following expression in which the operators are carefully ordered to avoid complications (c-numbers etc.) resulting from reordering.

$$P_m^{\ n} \equiv \sum_q \frac{(\ln a_m^\dagger)^q N_n^{\ q}}{q!} \frac{1}{\sqrt{N_n!}}$$

where the sum ranges from 0 to ∞. When $P_m^{\ n}$ is applied to a state it changes the value of the $m^{th}$ memory location from zero to the value at the $n^{th}$ memory location.

$$P_m^{\ n} | \ldots, \underset{m^{th}}{0}, \ldots, \underset{n^{th}}{x}, \ldots \rangle = \frac{(a_m^\dagger)^x}{\sqrt{x!}} | \ldots, 0, \ldots, x, \ldots \rangle$$

$$= | \ldots, x, \ldots, x, \ldots \rangle$$

The application of the factors of $N_n$ to the state results in factors of x that lead to the above expression when summed.

The operators $M_m$ and $P_m^{\ n}$ enable us to simply express the algebraic equivalent of assembly language instructions:

**LOAD m** – load the value at memory location m into the register

$$P_r^{\ m} M_r$$

**STORE m** – store the value in the register at memory location m

$$P_m^{\ r} M_m$$

**SHIFT k** – shift the value in the register by k bits. If k is positive the bit shift is to the right and if k is negative the bit shift is to the left. The bits are numbered from the leftmost bit which is bit 0 corresponding to $2^0$. The next bit is bit 1 corresponding to $2^1$ and so on.

If the bit shift is to the right (k > 0) then we assume the padding bits are 0's. For example a shift of the bit pattern for 7 = 1110000 … one bit to the right is 14 = 01110000 … As a result the value in the register is doubled (k = 1), quadrupled (k = 2), and so on. The algebraic expression for a k bit right shift is

$$\sum_q \frac{(\ln a_r^\dagger)^q S_r^{\ q}}{q!} \frac{\sqrt{N_r!}}{\sqrt{T_r!}}$$

where

$$S_r = (2^k - 1)N_r$$

and

$$T_r = 2^k N_r$$

If the bit shift is to the left (negative k), then we assume zero bits are added "at ∞". If k = -1 then the effect of left shift is to divide the value in the register by two (dropping the fractional part). If k = -2 then the effect of left shift is to divide the value in the register by four (dropping the fractional part) and so on. The algebraic expression that implements left shift is

$$\sum_q \frac{(\ln a_r)^q U_r^{\ q}}{q!} \frac{\sqrt{N_r!}}{\sqrt{V_r!}}$$

where

$$U_r = N_r - [2^k N_r]$$

and

$$V_r = [2^k N_r]$$

with [ z ] being the value of z truncated to an integer (fractional part dropped).

**ADD m** – add the value at memory location m to the value in the register

$$\sum_q \frac{(\ln a_r^\dagger)^q N_m^q}{q!} \frac{\sqrt{N_r!}}{\sqrt{(N_r + N_m)!}}$$

**SUBTRACT m** – subtract the value at memory location m from the value in the register (assumes the value in the register is greater than or equal to the value at location m)

$$\sum_q \frac{(\ln a_r)^q N_m^q}{q!} \frac{\sqrt{N_r!}}{\sqrt{(N_r - N_m)!}}$$

**MULTIPLY m** – multiply the value in the register by the value at memory location m

$$\sum_q \frac{(\ln a_r^\dagger)^q (N_r)^q (N_m - 1)^q}{q!} \frac{\sqrt{N_r!}}{\sqrt{(N_r N_m)!}}$$

**DIVIDE m** – divide the value in the register by the value at memory location m

$$\sum_q \frac{(\ln a_r^\dagger)^q W^q}{q!} \frac{\sqrt{N_r!}}{\sqrt{X!}}$$

where

$$W = N_r - [ N_r / N_m ]$$

and

$$X = [ N_r / N_m ]$$

with [ z ] being the value of z truncated to an integer (fractional part dropped).

**AND m** – change the value in the register by "and-ing" it with the value at memory location m

$$\sum_q \frac{((\ln a_r^\dagger)^q (W)^q \theta (W) + (\ln a_r)^q (-W)^q \theta (-W))}{q!} \frac{\sqrt{N_r!}}{\sqrt{X!}}$$

where

$$W = N_r \,\&\, N_m - N_r$$

and where

$$X = N_r \,\&\, N_m$$

with $\theta(z) = 1$ if $z > 0$ and 0 if $z < 0$. The & operator (adopted from the C programming language) performs bitwise AND. Corresponding bits in each operand are "multiplied" together using the multiplication rules:

$$1 \,\&\, 1 = 1$$

$$1 \,\&\, 0 = 0 \,\&\, 1 = 0 \,\&\, 0 = 0$$

For example the binary numbers 1010 & 1100 = 1000 or in base 10 5 & 3 = 1.

**OR m** – change the value in the register by "or-ing" it with the value at memory location m

$$\sum_q \frac{(\ln a_r^\dagger)^q (W)^q}{q!} \frac{\sqrt{N_r!}}{\sqrt{X!}}$$

where

$$W = N_r \,|\, N_m - N_r$$

and where

$$X = N_r \,|\, N_m$$

The | operator (adopted from the C programming language) performs bitwise OR. Corresponding bits in each operand are "multiplied" together using the multiplication rules:

$$1 \,|\, 1 = 1 \,|\, 0 = 0 \,|\, 1 = 1$$

$$0 \,|\, 0 = 0$$

For example the binary numbers $1010 \,|\, 1100 = 1110$ or in base 10 $5 \,|\, 3 = 7$.

**NOT** – change the value in the register by "noting" it

$$\sum_q \frac{((\ln a_r^\dagger)^q (W)^q \theta(W) + (\ln a_r)^q (-W)^q \theta(-W))}{q!} \frac{\sqrt{N_r!}}{\sqrt{X!}}$$

where

$$W = \sim\!N_r - N_r$$

and where

$$X = \sim\!N_r$$

with $\theta(z) = 1$ if $z > 0$ and 0 if $z < 0$. The ~ operator (adopted from the C programming language) performs bitwise NOT. Each 1 bit is replaced by a 0 bit and each 0 bit is replaced by a 1 bit. Since we have infinite words in our computer we supplement this rule with the restriction that the exchange of 1's and 0's only is made up to and including the rightmost 1 bit in the operand. The 0 bits beyond that remain 0 bits. For example the binary number $\sim\!101 = 010$ or in base 10, $\sim\!3 = 2$.

**INPUT m** – input a value storing it at memory location m. The input device is usually associated with a memory location from which the input symbolically takes place. We will designate the memory location of the input device as *in*.

$$P_m^{\,in} M_m$$

**OUTPUT m** – output the value at memory location m. The output device is usually associated with a memory location to which output symbolically takes place. We will designate the memory location of the output device as *out*.

$$P_{out}^{\,m} M_{out}$$

**TRA m** – set the value of the program counter to the value at memory location m. If we designate the program counter memory location as pc then this instruction is mapped to

$$P_{pc}^{\,m} M_{pc}$$

**TZR m** – set the value of the program counter to the value at memory location m if the value in the register is zero.

$$(P_{pc}^{\,m} M_{pc})^{\theta(N_r)\,\theta(-N_r)}$$

using $\theta(0) = 1$.

**HALT** – stop execution of the program. The halt in a program is mapped to a "bra" state vector.

$$< \ldots |$$

## A Simple Assembly Language Program

Assembly language instructions can be combined to form an assembly language program. Perhaps the best way to see how the algebraic representation of assembly language works is to translate a simple assembly language program into its algebraic equivalent.

The program that we will consider is:

1  INPUT x
2  INPUT y
3  LOAD x
4  ADD y
5  STORE z
6  OUTPUT z
7  HALT

This program translates to the algebraic equivalent:

$$\underset{7}{< \ldots | P_{out}^z M_{out}} \;\; \underset{6}{P_z^r M_z} \;\; \underset{5}{\frac{(a_r^\dagger)^{N_y} \sqrt{N_r!}}{\sqrt{(N_r + N_y)!}}} \;\; \underset{4}{P_r^x M_r} \;\; \underset{3}{P_y^{in} M_y} \;\; \underset{2}{P_x^{in} M_x | \ldots >} \;\; \underset{1}{\text{Steps}}$$

where the power of $a_r^\dagger$ is represented by a power series expansion as seen earlier.

The algebraic expression in the brackets produces one output state from a given initial state. The values in memory after the last step correspond to one and only one output state of the form:

$$< n, m, p, \ldots | = (N (r^\dagger)^n (a_0^\dagger)^m (a_1^\dagger)^p \ldots | 0 >)^\dagger$$

where N is a normalization constant.

This simple program does not produce a superposition of states. As a result programs of this type are analogous to ordinary programs for normal, non-Quantum computers. The numbers in memory after the program concludes are the "output" of the program. We will see programs in succeeding sections that take a computer of fixed state N $(r^\dagger)^n (a_0^\dagger)^m (a_1^\dagger)^p \ldots | 0 >$ and produce a superposition of states that must be interpreted quantum mechanically. These programs are quantum in nature and the computer that runs them must be a quantum computer.

## Programs and Program Logic

The simple program of the last section corresponded to a sequential program that executed step by step. We now turn to more complex programs with program logic that supports non-sequential execution of programs. When this type of program executes the execution of the instructions can lead to jumps from one instruction to another instruction in another part of the program.

Programs are linear – one instruction executes after another. But they are not sequential – the instructions do not always execute step by step sequentially. A program can specify jumps ("goto" instructions) in the code from the current instruction to an instruction several steps after the current instruction or several steps back to a previous instruction. The code then executes sequentially until the next jump is encountered.

These jumps in the code at the level of assembly language implement the control constructs such as goto statements, if expressions, for loops, and switch expressions seen in higher level languages such as C and C++.

Jumps in code can be implemented in the algebraic representation of programs by having a program counter memory value that increments as the algebraic factor corresponding to each step executes. Steps in the program can execute or not execute depending on the current value of the program counter.

Changes in the program counter value are made using the TRA and TZR instructions. In the algebraic representation the program counter variable can be used to manage the execution of the program steps.

The key algebraic constructs supporting non-sequential program execution are:

Execute instruction only if PC ≤ n

$$( \ldots )^{\theta ( n - N_{pc})}$$

Execute instruction only if PC ≥ n

$$( \ldots )^{\theta ( N_{pc} - n)}$$

Execute instruction only if PC = n

$$( \ldots )^{\theta\theta ( N_{pc} - n)}$$

Execute instruction only if PC not equal to n

$$( \ldots )^{\theta( N_{pc} - n) + \theta( n - N_{pc}) - 2\theta\theta( N_{pc} - n)}$$

where the parentheses contain one or more instructions and where $\theta\theta(x) = 1$ if $x = 0$ and zero otherwise. The function $\theta\theta(x)$ can be represented by step functions as

$$\theta\theta(x) = \theta(x) \theta(-x)$$

Using these constructs we can construct non-sequential programs that supprt "goto's", if's and other control constructs seen in higher level languages.

To illustrate this feature of the algebraic representation we will consider an enhancement of the assembly language program seen earlier:

```
1    INPUT x
2    INPUT y
3    LOAD x
4    TZR y
5    ADD y
6    STORE z
7    OUTPUT z
8    HALT
```

This program has the new feature that if the first input – to memory location x – is zero, then instruction 4 will cause a jump to the instruction specified by the value stored at memory location y.

For example if the inputs are 0 placed at memory location x and 2 placed at memory location y, then the TZR instruction will cause the program to jump to instruction 2 from instruction 4. Then the program will proceed to execute from instruction 2.

Another example of a case with a jump is if the input to memory location x is zero and the input to memory location y is 6 then the program jumps from instruction 4 to instruction 6 and the program completes execution from there. If the input to memory location x is non-zero no jump takes place.

To establish the algebraic equivalent of the preceding example we have to use the non-sequential constructs provided earlier in this section. In addition we must define the equivalent recursively because of the possibility that the program may jump backwards to an earlier instruction in the program. If only "forward" jumps were allowed then recursion would not be needed.

An algebraic representation of the program that supports only forward leaps is:

$$\overset{8}{<} \ldots | \, ( a_{pc}^{\dagger} P_{out}^{\, z} M_{out})^{\theta\theta(N_{pc}- 7)}$$

$$(a_{pc}^\dagger P_z^r M_z)^{\theta\theta(N_{pc}-6)}$$

$$\left(\frac{a_{pc}^\dagger (a_r^\dagger)^{N_y} \sqrt{N_r!}}{\sqrt{(N_r + N_y)!}}\right)^{\theta\theta(N_{pc}-5)}$$

$$(a_{pc}^\dagger)^{1-\theta\theta(N_r)} (P_{pc}^y M_{pc})^{\theta\theta(N_r)\,\theta\theta(N_{pc}-4)}$$

$$\overset{3}{a_{pc}^\dagger P_r^x M_r}$$

$$\overset{2}{a_{pc}^\dagger P_y^{in} M_y}$$

$$\overset{1}{a_{pc}^\dagger P_x^{in} M_x}$$

$$a_{pc}^\dagger M_{pc} | \ldots >$$

The program steps are numbered above each corresponding expression. The step function expressions enable the jump to take place successfully.

A program with forward and backward jumps supported requires a recursive definition. We will define the recursive function f() with:

$$f() = (a_{pc}^\dagger P_{out}^z M_{out})^{\theta\theta(N_{pc}-7)}$$

$$(a_{pc}^\dagger P_z^r M_z)^{\theta\theta(N_{pc}-6)}$$

$$\left(\frac{a_{pc}^\dagger (a_r^\dagger)^{N_y} \sqrt{N_r!}}{\sqrt{(N_r + N_y)!}}\right)^{\theta\theta(N_{pc}-5)}$$

$$(a_{pc}^\dagger)^{1-\theta\theta(N_r)} (f() P_{pc}^y M_{pc})^{\theta\theta(N_r)\,\theta\theta(N_{pc}-4)}$$

$$(\,a_{pc}^\dagger P_r^x M_r\,)^{\theta\theta(N_{pc}-3)}$$

$$(\,a_{pc}^\dagger P_y^{in} M_y\,)^{\theta\theta(N_{pc}-2)}$$

$$(\,a_{pc}^\dagger P_x^{in} M_x\,)^{\theta\theta(N_{pc}-1)}$$

The program is

$$f() a_{pc}^\dagger M_{pc} | \ldots >$$

This program is well behaved except if the input value placed at the y memory location is 4. In this case the program recursively executes forever. This defect can be removed by using another memory location for a counter variable.

We can modify the program so that the program only recursively calls itself a finite number of times by having each recursive call decrease the counter variable by one. When the value reaches zero the recursion terminates. An example of such a program (set to allow at most 10 iterations of the recursion) is:

$$g() = (a_{pc}^\dagger P_{out}^z M_{out})^{\theta\theta(N_{pc}-7)}$$

$$(a_{pc}^\dagger P_z^{\ r}\ M_z)^{\theta\theta(N_{pc}-6)}$$

$$\left(\frac{a_{pc}^\dagger\ (a_r^\dagger)^{N_y}\ \sqrt{N_r!}}{\sqrt{(N_r+N_y)!}}\right)^{\theta\theta(N_{pc}-5)}$$

$$(a_{pc}^\dagger)^{1-\theta\theta(N_r)}\ ((a_{pc}^\dagger)^{1-\theta(N_w)}(g())^{\theta(N_w)}a_w P_{pc}^{\ y}\ M_{pc})^{\theta\theta(N_r)\ \theta\theta(N_{pc}-4)}$$

$$(\ a_{pc}^\dagger P_r^{\ x} M_r\ )^{\theta\theta(N_{pc}-3)}$$

$$(\ a_{pc}^\dagger P_y^{\ in}\ M_y)^{\theta\theta(N_{pc}-2)}$$

$$(\ a_{pc}^\dagger P_x^{\ in}\ M_x)^{\theta\theta(N_{pc}-1)}$$

The program is

$$g()a_{pc}^\dagger M_{pc}\ (a_w^\dagger)^{10} M_w\ |\ \ldots >$$

where w is some memory location. We conjecture that any assembly language program using the previously specified instructions can be mapped to an algebraic representation – possibly with the use of additional memory for variables such as the counter variable seen above.

Using the algebraic constructs supporting non-sequential program execution we can create algebraic representations of assembly language programs. These programs have a definite input state and through the execution of the program they evolve into a definite output state -–not a superposition of output states. Therefore they faithfully represent assembly language programs. On the other hand they are quantum in the sense that they use states and harmonic oscillator-like raising and lowering operators. The types of programs we are creating in this approach are "sharp" on the space of states. One input state evolves through the program's execution to one and only one output state with probability one.

These types of programs are analogous to free field theory in which incoming particles evolve without interaction to an output state containing the same particles.

In the next section we extend the ideas in this section to quantum programming where a variety of output states are possible – each with a certain probability of being produced.

**Quantum Assembly Language™ Programs**

In this section we will first look at a simplified quantum program that illustrates quantum effects but in actuality is a sum of deterministic assembly language programs mapped to algebraic equivalents. Consider a "quantum" program that is the sum of three ordinary programs $g_1()$, $g_2()$ and $g_3()$ of the type seen in the last section. Further let us assume the set of orthonormal states

$$|\ n, m, p, \ldots >$$

that we saw in the previous sections with

$$< X\ |\ Y > = \delta_{XY}$$

where $\delta_{XY}$ represents a product of Kronecker $\delta$ functions in the individual values in memory of the $|\ X >$ and $|\ Y >$ states. Further let us assume

$$|\ n_1, m_1, p_1, \ldots > = g_1()|\ \ldots >$$

$$|\ n_2, m_2, p_2, \ldots > = g_2()|\ \ldots >$$

$$| n_3, m_3, p_3, \ldots > = g_3()| \ldots >$$

for some initial state of the quantum computer. Then

$$\alpha g_1() + \beta g_2() + \gamma g_3()| \ldots >$$

is a "quantum" program where $\alpha$, $\beta$, and $\gamma$ are constants such that

$$|\alpha|^2 + |\beta|^2 + |\gamma|^2 = 1$$

The quantum program produces the state $| n_1, m_1, p_1, \ldots >$ with probability $|\alpha|^2$, the state $| n_2, m_2, p_2, \ldots >$ with probability $|\beta|^2$, and the state $| n_3, m_3, p_3, \ldots >$ with probability $|\gamma|^2$.

We now have a quantum probabilistic computer. The programs $g_1()$, $g_2()$ and $g_3()$ are being executed in *parallel* in a quantum probabilistic manner.

Currently, the most feasible way of creating a Quantum Computer with current technology or reasonable extrapolations of current technology is to create a material which approximates a lattice with spins at each lattice site that we can orient electromagnetically at the beginning of a program. The execution of a program takes place by applying electromagnetic fields that have a time dependence specific to the computation. The electromagnetic fields implement a custom-tailored set of interactions between the spins in the material that simulates the calculation to be performed.

The interactions are specified with some Hamiltonian or some effective Hamiltonian and the initial state of the lattice spins evolves dynamically to some configuration that is then measured.

The Hamiltonians are normally specified using the space-time formalism that is a familiar part of Quantum Mechanics. A Hamiltonian specifies the time evolution of a system starting from an initial state. We can introduce an explicit time dependence in states by using the notation:

$$| \Psi(t) >$$

to denote the state of a Quantum Computer at time t. The general state of the computer at time t can be written as a superposition of the number representation states:

$$| \Psi(t) > = \sum_n f_n(t)| n_1, n_2, n_3, \ldots >$$

where n represents a set of values $n_1, n_2, n_3, \ldots$

The time evolution of the states can be specified using the Hamiltonian operator H as

$$| \Psi(t) > = e^{-iHt}| \Psi(0) >$$

With this Hamiltonian formulation we can imagine wishing to simulate a physical (or mathematical) process, defining a Hamiltonian that corresponds to the process, and then creating an experimental setup using a set of lattice spins in some material that implements the simulation. The experimental setup will prepare the initial state of the spins, create a fine tuned interaction that simulates the physics of the process, and then, after the system has evolved, will measure the state of the system at time t. Repeated performance of this procedure will determine the probability distribution associated with the final state of the Quantum Computer. The probability distribution is specified by $|f_n(t)|^2$ as a function of the sets of numbers denoted by n.

A simple example of a Hamiltonian that causes a Quantum Computer to evolve in a non-trivial way is:

$$H = \sum_{m=0}^{\infty} a_{m+1}^{\dagger} a_m$$

(This example was chosen partly because it has a form similar to a Virasoro algebra generator in SuperString Theory.) Let us assume the initial state of the Quantum Computer at t = 0 is

$$| 1, 0, 0, 0, \ldots >$$

that is, an initial value of 1 in the first word in memory and zeroes in all other memory locations. At time t the state of memory is:

$$| \Psi(t) > = \sum_{n=0}^{\infty} f_n(t) | \underset{\underset{n^{th} \text{ memory location}}{\uparrow}}{0, 0, \ldots, 1, 0, \ldots} >$$

with

$$f_n(t) = (-it)^n/n!$$

using the power series expansion of the exponentiated Hamiltonian expression. The probability of finding the state

$$| \underset{\underset{n^{th} \text{ memory location}}{\uparrow}}{0, 0, \ldots, 1, 0, \ldots} >$$

is

$$(t^n/n!)^2$$

At first glance the Hamiltonian approach is very different from the Quantum Assembly Language™ approach discussed above. However these approaches can be interrelated in special cases and (we conjecture) in the general case through sufficiently clever transformations. For example, the preceding Hamiltonian can be re-expressed as assembly language instructions

$$H = \sum_{m=0}^{\infty} (\text{STORE } (m+1))(\text{ADD “1”})(\text{LOAD } (m+1)) \cdot$$

$$\cdot (\text{STORE } m)(\text{SUBTRACT “1”})(\text{LOAD } m)$$

where a value is loaded into the register from memory location m and then 1 is added to the value in the register. The "1" expression represents a literal value one not a memory location. The parentheses around m+1 indicates it is the $(m+1)^{th}$ memory location – not the addition of one to the value at the $m^{th}$ location.

The preceding assembly language expression for H can be replaced with the algebraic representation expression:

$$H = \sum_{m=0}^{\infty} P_{m+1}^{r} M_{m+1} a_r^{\dagger} \frac{1}{\sqrt{(N_r + 1)}} P_r^{m+1} M_r P_m^{r} M_m a_r \sqrt{N_r} P_r^{m} M_r$$

This complex expression is not an improvement in one sense. The original Hamiltonian expression was much simpler. Its importance is the mapping that it embodies from a quantum mechanical Hamiltonian to an assembly language expression to an algebraic representation of the assembly language.

If we regard the value in the register as a "scratchpad" value as programmers often do, then we can establish a representation of $a_m^{\dagger}$ and $a_m$ in terms of the algebraic representation of assembly language instructions.

$$a_m^{\dagger} \equiv P_m^{r} M_m a_r^{\dagger} \frac{1}{\sqrt{(N_r + 1)}} P_r^{m} M_r$$

and

$$a_m \equiv P_m^{r} M_m a_r \sqrt{N_r} P_r^{m} M_r$$

The power series expansion of the exponentiated Hamiltonian in the previous example is an example of the use of Perturbation Theory. The direct solution of a problem is often not feasible because of the complexity of the dynamics. Physicists have a very well developed theory for the approximate solution of these difficult problems called Perturbation Theory. Perturbation Theory takes an exact solution of a simplified version of the problem and then calculates corrections to that solution that approximate the exact solution of the problem. In the preceding example the initial state of the Quantum Computer represents a time-independent description of the Quantum Computer. The time-dependent description of the Quantum Computer which is the sought-for solution requires the evaluation of the result of the application of the exponentiated Hamiltonian to the initial state. For a small elapsed time, the exponential can be expanded in a power series and the application of the first few terms of the power series to the initial state approximates the actual state of the Quantum Computer. Thus we have a Perturbation Theory for the time evolution of the Quantum Computer expressed as an expansion in powers of the elapsed time.

**Bit-Level Quantum Computer Language**
In the previous section we examined a Quantum Assembly Language™ with words consisting of an infinite sets of bits. In this section we will examine the opposite extreme – a Quantum Computer Language with one-bit words. One can also create Quantum Computer Languages for intermediate cases such as 32-bit words.

A Bit-Level Quantum Computer Language can be represented with anti-commuting Fermi operators $b_i$ and $b_i^\dagger$ for i = 0, 1, 2, … representing each bit location in the Quantum Computer's memory with the anti-commutation rules:

$$\{ b_i, b_j^\dagger \} = \delta_{ij}$$

$$\{ b_i, b_j \} = 0$$

$$\{ b_i^\dagger, b_j^\dagger \} = 0$$

where $\delta_{ij}$ is 1 if i = j and zero otherwise. We will assume an (unrealistic) one-bit register with a pair of raising and lowering operators r and $r^\dagger$ for the register with the anti-commutation relations:

$$\{ r_i, r_j^\dagger \} = \delta_{ij}$$

$$\{ r_i, r_j \} = 0$$

$$\{ r_i^\dagger, r_j^\dagger \} = 0$$

The ground state of the computer is the state with the values at all bit memory locations set to zero. It is represented by the vector

$$| 0, 0, 0, \ldots > \equiv | 0 > \equiv \Phi_V$$

A typical state of the computer will be represented with a vector such as

$$| 1, 1, 1, \ldots > \; = \; r^\dagger b_0^\dagger b_1^\dagger \ldots | 0 >$$

with the first number being the value in the register, the second number the value at memory location 0, the third number the value at memory location 1, and so on.

A specified Quantum Computer state evolves as a Quantum Computer Program executes to a final computer state. A Bit-Level Quantum Computer Program can be represented as an algebraic expression in anti-commuting raising and lowering operators. The approach is similar to the approach seen earlier in this chapter for infinite-bit words using commuting operators.

**Basic Operators of the Bit-Level Quantum Language**
The key operators that are required for the algebraic representation of a Bit-Level Quantum Computer Language™ are:

Fetch the Value at a Memory Location (Number Operator)

$$N_m = b_m^\dagger b_m$$

For example,

$$N_m | \ldots, 1, \ldots \rangle = | \ldots, 1, \ldots \rangle$$

↑ $m^{th}$ memory location value

Set the Value at Memory Location m to Zero

$$M_m = (b_m)^{N_m}$$

The above expression for $M_m$ is symbolic. The expression represents the following expression in which the operators are carefully ordered to avoid complications (c-numbers etc.) resulting from reordering.

$$M_m \equiv e^{N_m \ln b_m} = \sum_q \frac{(\ln b_m)^q N_m^q}{q!}$$

where the sum ranges from 0 to ∞. $M_m$ becomes

$$M_m = 1 + (b_m - 1)N_m$$

using the identity $N_m = N_m^2$. When $M_m$ is applied to a state it sets the value of the $m^{th}$ memory location to zero.

$$M_m | \ldots, x, \ldots \rangle = | \ldots, 0, \ldots \rangle$$

↑ $m^{th}$ memory location value

Change the Value at Memory Location m from 0 to the Value at Location n

$$P_m^n = (b_m^\dagger)^{N_n}$$

The above expression for $P_m^n$ is also symbolic. The expression represents the following expression in which the operators are carefully ordered to avoid complications (c-numbers etc.) resulting from reordering.

$$P_m^n \equiv \sum_q \frac{(\ln b_m^\dagger)^q N_n^q}{q!}$$

where the sum over q ranges from 0 to ∞. Using the identity $N_m = N_m^2$ the expression for $P_m^n$ simplifies to:

$$P_m^n = 1 + (b_m - 1)N_m$$

When $P_m^n$ is applied to a state it changes the value of the $m^{th}$ memory location from zero to the value at the $n^{th}$ memory location.

$m^{th}$ ↓ $n^{th}$ ↓

$$P_m^n | \ldots, 0, \ldots, x, \ldots \rangle = (b_m^\dagger)^x | \ldots, 0, \ldots, x, \ldots \rangle$$

$$= | \ldots, x, \ldots, x, \ldots \rangle$$

We can use the operators $M_m$ and $P_m^n$ to express bit-wise assembly language instructions:

**LOAD m** – load the value at memory location m into the register

$$P_r^m M_r = (1 - N_r + b_r)(1 - N_m) + (N_r + b_r^\dagger)N_m$$

The first term on the right handles the case $N_m = 0$ and the second term on the right handles the case $N_m = 1$.

**STORE m** – store the value in the register at memory location m

$$P_m^r M_m = (1 - N_m + b_m)(1 - N_r) + (N_m + b_m^\dagger)N_r$$

The first term on the right handles the case $N_r = 0$ and the second term on the right handles the case $N_r = 1$.

**ADD m** – add the value at memory location m to the value in the register

$$(b_r^\dagger)^{N_m} = \sum_q \frac{(\ln b_r^\dagger)^q N_m^q}{q!}$$

$$= 1 + (b_r^\dagger - 1)N_m$$

If both the register and memory bit m have values of one then the application of this operator expression to the quantum state produces zero.

**SUBTRACT m** – subtract the value at memory location m from the value in the register

$$(b_r)^{N_m} = \sum_q \frac{(\ln b_r)^q N_m^q}{q!}$$

$$= 1 + (b_r - 1)N_m$$

If the value in the register is zero and the value at location m is one the application of this operator produces zero.

**MULTIPLY m** – multiply the value in the register by the value at memory location m

$$(b_r^\dagger)^{(N_m-1)N_r} = \sum_q \frac{(\ln b_r^\dagger)^q (N_r)^q (N_m - 1)^q}{q!}$$

$$= 1 + (b_r - N_r)(1 - N_m)$$

Other assembly language instructions can be expressed in algebraic form as well.

The operator algebra that we have developed for a bit-wise Quantum Assembly Language™ or a Quantum Machine Language™ provides a framework for the investigation of the properties of Quantum Languages within an algebraic framework – a far simpler task than the standard quantum linguistic approaches.

## Quantum High Level Computer Language Programs

The Quantum Assembly Language™ representation that we have developed earlier in this chapter forms a basis for high level Quantum Programming Languages. These languages are analogous to high level computer languages such as C or C++ or FORTRAN.

In ordinary computation a statement in a high level language such as

$$a = b + c;$$

in C programming is mapped to a set of assembly language by a C compiler. A simple mapping of the above C statement to assembly language would be

```
LOAD ab
ADD  ac
STORE aa
```

where aa is the memory address of a, ab is the memory address of b and ac is the memory address of c.

If we decide to define a High Level Quantum Computer Language™ then it would be natural to define it analogously in terms of a Quantum Assembly Language™. A statement in the High Level Quantum Computer Language™ would map to a set of Quantum Assembly Language™ instructions.

For example, a = b + c would map to the algebraic expression

$$P_{aa}{}^r \, M_{aa} \; (a_r{}^\dagger)^{N_{ac}} \, \frac{\sqrt{N_r!}}{\sqrt{(N_r + N_{ac})!}} \, P_r{}^{ab} \, M_r$$

using the formalism developed earlier in this chapter to LOAD, ADD and STORE.

The definition of high level Quantum Computer Languages™ in this approach is straightforward. One can then imagine creating programs in these languages for execution on Quantum Computers just as ordinary programs are created for ordinary computers.

Another approach to higher level Quantum Computer Languages™ is to simply express them directly using raising and lowering operators – not in terms of Quantum Assembly Language™ instructions. For example the preceding a = b + c; statement can be directly expressed as

$$(a_{aa}{}^\dagger)^{N_{ac}+N_{ab}} \, (a_{aa})^{N_{aa}} \, \frac{1}{\sqrt{(N_{ac} + N_{ab})!} \, \sqrt{N_{aa}!}}$$

Simple High Level Quantum Computer programs can be expressed as products of algebraic expressions embodying the statements of the program. These programs are sharp on the set of memory states taking an initial memory state that is an eigenstate of the set of number operators $N_m$ into an output eigenstate of the number operators.

A general High Level Quantum Computer Program is a sum of simple High Level Programs. For example,

$$\alpha h_1() + \beta h_2() + \gamma h_3() | \ldots >$$

where $\alpha$, $\beta$, and $\gamma$ are constants such that

$$|\alpha|^2 + |\beta|^2 + |\gamma|^2 = 1$$

The sum of simple programs $\alpha h_1() + \beta h_2() + \gamma h_3()$ produces the state $| n_1, m_1, p_1, \ldots >$ with probability $|\alpha|^2$, the state $| n_2, m_2, p_2, \ldots >$ with probability $|\beta|^2$, and the state $| n_3, m_3, p_3, \ldots >$ with probability $|\gamma|^2$.

An initial eigenstate of the number operators is tranformed into an output state that is a superposition of number operator eigenstates. In this case we use probabilities to specify the likelihood that a given output eigenstate will be found when the output state is measured.

A Hamiltonian can also be used to specify the time evolution of a system starting from an initial state. Using the notation:

$$| \Psi(t) >$$

to denote the state of a Quantum Computer at time t the general state of a computer at time t can be written as a superposition of number representation states:

$$| \Psi(t) > = \sum_n f_n(t) | n_1, n_2, n_3, \ldots >$$

where n represents a set of values $n_1, n_2, n_3, \ldots$

The time evolution of the states can be specified using the Hamiltonian operator H as

$$| \Psi(t) > = e^{-iHt} | \Psi(0) >$$

A simple example of a Hamiltonian that causes a Quantum Computer to evolve in a non-trivial way is:

$$H = \sum_{m=0}^{\infty} (a_{m+2}^{\dagger})^{N_{m+1}+N_m} (a_{m+2})^{N_{m+2}} \frac{1}{\sqrt{(N_{m+1}+N_m)!} \sqrt{N_{m+2}!}}$$

This Hamiltonian is based on the a = b + c statement above. This Hamiltonian generates a complex superposition of states as time evolves. More complex Hamiltonians equivalent to programs with several statements can be easily constructed.

## Quantum C Language™

One of the most important computer languages is the C programming language developed at Bell Laboratories in the 1970's. The original version of version of the C language was a remarkable combination of low level (assembly language-like) features and high level features like the mathematical parts of FORTRAN. The variables in the language were integers stored in words just as we saw in the earlier examples in this chapter. (There were several other types of integers as well – a complication that we will ignore.)

Using the ideas seen in the earlier sections of this chapter it is easy to develop algebraic equivalents for most of the constructs of the C language and thus create a Quantum C Language™. An important element that must be added to the previous development is to introduce the equivalent of pointers. Simply put pointers are variables that have the addresses of memory locations as their values. The C language has two important operators for pointer manipulations:

| Operator | Role | Example |
|---|---|---|
| & | Fetch an address | ptr = &x; |
| * | Fetch/set the value at an address | z = *ptr; <br> *ptr = 99; |

The & operator of C fetches the address of a variable in memory. The example shows a pointer variable ptr being set equal to the address of the x variable. The * (dereferencing) operator can fetch the value at a memory location. The first * example illustrates this aspect: the variable z is set equal to the value at the memory location specified by the pointer variable ptr. The * operator can also be used to set the value at a memory location as illustrated by the second * example. In this example the value 99 is placed at the memory location (address) specified by the ptr pointer variable.

These operators can be implemented in the algebraic representation of the Quantum C Language™ in the following way:

$$\& \quad \Leftrightarrow \quad [A, ]$$

where $A = \Sigma\, m(a_m - a_m^{\dagger})$ with the sum from 0 to $\infty$. If we apply the operator to a raising or lowering operator we obtain its address

$$\& \, a_m^{\dagger} = [A, a_m^{\dagger}] = m = \& \, a_m = [A, a_m]$$

The equivalent of the * operator is actually a pair of operator expressions. To fetch the value at a memory location we use

$$*m \equiv N_m$$

To set the value to X at a memory location m we use a more complex C language representation:

$$*m = X;$$

An equivalent algebraic expression is:

$$*m(X) \equiv (a_m^{\dagger})^X (a_m)^{N_m} \frac{1}{\sqrt{X!} \sqrt{N_m!}}$$

*m(X) is a functional notation. So a = b + c can be rewritten as a "pointer" algebraic expression as:

$$(a_{aa}^{\dagger})^{*ac + *ab} (a_{aa})^{*aa} \frac{1}{\sqrt{(*ac + *ab)!} \sqrt{*aa!}}$$

Or more compactly using the functional notation as

$$*aa(*ab + *ac)$$

The Quantum C Language™ could be used to define Hamiltonians for a Quantum Computer. Other languages such as Java™, C++, lisp and so on also have Quantum analogues which may be defined in a similar way.

**SuperString Quantum Computer Revisited**
The concepts we have been exploring for the algebraic representation of Quantum Programs raise the question - Can the SuperString Quantum Computer and Superstring interactions in particular be cast into the form of algebraic quantum computations.

It appears that it may be possible to accomplish this purpose. The basis for this formulation is first the use of the raising and lowering operator formalism as the primary formalism and the relegation of the space-time formalism of SuperStrings to a secondary status. Secondly, the state-operator isomorphism of SuperString theory can be used to develop a representation of vertex operators for interactions as as states within the processor part of the SuperString Quantum Computer. This surprising feature of SuperString theory answers a very basic question confronting the SuperString Quantum Computer: How do we represent SuperString interactions in the Quantum Computer framework. The answer appears to be that we treat interactions (vertex operators) as states within the Quantum Computer. Without this feature of SuperString theory the representation of interactions in the SuperString Quantum Computer would be an open question.